%% file: p00_main.tex
\documentclass[10pt]{article}
\pdfoutput=1
\usepackage{amsmath}
\usepackage{tabularx}
\usepackage{lscape}
\usepackage{afterpage}
\usepackage{lineno}
\usepackage[a4paper]{geometry}
\usepackage{textgreek}
\usepackage[utf8]{inputenc}
\usepackage{graphicx}
\usepackage{tikz}
\usepackage[sorting=none,style=nature]{biblatex}
\usepackage[symbol]{footmisc}
\usepackage{nameref}
\usetikzlibrary{shapes,arrows,positioning,decorations.pathreplacing}
\title{Quantitative model for \textsuperscript{13}C tracing applied to citrate production and secretion of prostate epithelial tissue}
\author{Frits H.A. van Heijster\textsuperscript{1}, Vincent Breukels\textsuperscript{1}, Arend Heerschap\textsuperscript{1}}
\date{\small{\textsuperscript{1}Radiology and Nuclear Medicine, Radboud University Medical Center, Nijmegen, The Netherlands}}

\begin{document}
\maketitle
\noindent

\input{p00_abstract}
\newpage
\input{p01_Introduction.tex}
\input{p02_Model.tex}
\input{p03_Derivation_glc_16.tex}
\input{p04_Derivation_pyr_2.tex}
\input{p05_Model.tex}
\input{p06_Calculations.tex}
\input{p10_Table_2.tex}
\input{p07_DiscussionConclusion.tex}
\input{p08_Appendices.tex}
\printbibliography
\end{document}

%% file: p00_abstract.tex
\section{Abstract}
Healthy human prostate epithelial cells have the unique ability to produce and secrete large amounts of citrate into the lumen of the prostate. Citrate is a Krebs cycle metabolite produced in the condensation reaction between acetyl-CoA and oxaloacetate in the mitochondria of the cell. With the application of \textsuperscript{13}C enriched substrates, such as \textsuperscript{13}C glucose or pyruvate, to prostate cells or tissues, it is possible to identify the contributions of different metabolic pathways to this production and secretion of citrate. In this work we present a quantitative model describing the mitochondrial production and the secretion of citrate by prostatic epithelial cells employing the \textsuperscript{13}C labeling pattern of secreted citrate as readout. We derived equations for the secretion fraction of citrate and the contribution of pyruvate dehydrogenase complex versus the anaplerotic pyruvate carboxylase pathways in supplying the Krebs cycle with carbons from pyruvate for the production of citrate. These measures are independent of initial \textsuperscript{13}C-enrichment of the administered supplements and of \textsuperscript{13}C J-coupling patterns, making this method robust also if SNR is low. We propose the use of readout variables of this model to distinguish between citrate metabolism in healthy and diseased prostate tissue, in particular upon malignant transformation.

%% file: p01_Introduction.tex
\section{Introduction}
Healthy prostate epithelial cells have the unique capability of secreting citrate in the ducts or lumen of the prostate.\supercite{Costello1997}, where it can reach levels up to about 180 mM. Citrate accumulation in the prostate is promoted by inhibition of the citrate converting enzyme \textit{m}-aconitase through zinc binding, which is taken up at relatively high levels in epithelial prostate cells.\supercite{Costello1997,Costello1999,Bertilsson2012} The characteristic citrate accumulation is lost upon malignancy,\supercite{Singh2006} and a decreased citrate signal in MR spectroscopic images of the prostate is used as a biomarker for the presence of cancer.\supercite{Heerschap,Kurhanewicz1995,Kurhanewicz2016} Prostate cancer accounts for one in five cancers and is responsible for one in ten cancer related deaths in men worldwide.\supercite{Bray2013,Ferlay2013} Understanding metabolic reprogramming in malignant transformation may help to better diagnose and treat prostate cancer.\supercite{Kelly2016,Tayari2017,Testa2016,Twum-ampofo2016} 
Citrate is the product of a condensation reaction of oxaloacetate and acetyl-CoA catalyzed by citrate synthase as the first step of the Krebs cycle. A major compound from which acetyl-CoA is derived is glucose, which is metabolized into pyruvate by glycolysis and subsequently enters the the mitochondria where it is converted into acetyl-CoA to enter the Krebs cycle. In this step two carbons enter the cycle, which is equal to the carbon efflux as carbon dioxide, during one Krebs cycle turn. Because of citrate efflux into the luminal space, removing six carbons from the cycle, anaplerotic contributions are required to maintain the carbon pool of the Krebs cycle. One of these anaplerotic contributions can come from pyruvate being converted into oxaloacetate by pyruvate carboxylase (PC). The influx of this four-carbon metabolite can partly balance the efflux of six carbon citrate every Krebs cycle turn.
\bigskip\\
The carbon flow for citrate production and secretion can be assessed by supplying prostate tissue or cells \textit{in vivo} or \textit{in vitro} with \textsuperscript{13}C substrates and monitoring the fate of the \textsuperscript{13}C-labels by \textsuperscript{13}C MR spectroscopy or high resolution \textsuperscript{13}C NMR spectroscopy and subsequently looking at the specific \textsuperscript{13}C labeling of metabolic products. In this way metabolic routes involved in citrate generation can be delineated and quantified.
\bigskip\\
Multiple metabolic flux modeling studies have modeled Krebs cycle activity in different tissues and cell types under different conditions. This has been done using (extensive) metabolic networks involving fluxes between the metabolites in the network, like positional isotopomers\supercite{Schmidt1997}, bonded cumomer analysis\supercite{Wiechert1999} or elementary metabolite unit based methods\supercite{Antoniewicz2007}. For example, in human melanoma cell line DB1, rat C6 glioma cells, H460 NSCLC cells, hepatocytes and brown adipocytes and in isolated cardiac mitochondria relative fluxes of total TCA cycle activity $V_{TCA}$ and exchange flux $V_x$ of $\alpha$-ketoglutarate with glutamate $V_x$ have been determined \textit{in vitro}.\supercite{Shestov2016,Shestov2016a,PORTAIS1993,Wu2007,Jiang2017,Young2008,Egnatchik2018} But also \textit{in vivo} in mouse and rat brain and heart \supercite{Xin2015,Carvalho2004,Malloy1996,Burgess2001,Yang2006} and human brain.\supercite{Chen2001,Mason1995,Cheshkov2017} The fast exchange between glutamate and $\alpha$-ketoglutarate is one common element in these studies. Overall, the different contributions of pyruvate carboxylase activity, malic enzyme activity, glutaminolysis and other cataplerotic and anaplerotic processes\supercite{Owen2002} for different tissues and cells make it hard to generalize Krebs cycle metabolism in such a way that it can be used to predict prostate metabolism. Moreover, the unique features of citrate metabolism in the prostate make it hard to translate values found in these other studies to prostate metabolism. Most NMR investigations with \textsuperscript{13}C labeling of prostate tissue have been performed with hyperpolarization of the substrates\supercite{Chen2017,BancroftBrown2019}, but none have been focusing on citrate metabolism.
\bigskip\\
The unique characteristic secretion of citrate by prostatic cells provides a six-carbon readout of this Krebs cycle activity, which has not been explored before as far as we know. A first practical approach to explore this property would be in studies of prostate cells, but most prostatic epithelial cell lines lack the capacity of citrate secretion. However, the prostate cancer metastasis cell line LNCaP has been shown to still have the capacity to secrete citrate and therefore may serve as a convenient model system to evaluate if \textsuperscript{13}C labeling of secreted citrate can be explored as a readout for carbon flow inside the cell.\supercite{Cornel1995,Franklin1995,Korenchuk} Using only simple ratios of the presence of \textsuperscript{13}C in citrate, which can be extracted from MR spectroscopic data or high resolution \textsuperscript{13}C NMR spectra, we developed a simple quantitative model for cellular citrate production in this work. On one hand we determine an index for citrate production and secretion versus Krebs cycle consumption and on the other hand an estimate of the balance between pyvuvate carboxylase (PC) and pyruvate dehydrogenase complex (PDC) activity in citrate production.

%% file: p02_Model.tex
\section{Scope of data suitable for the model}
To follow citrate metabolism in healthy or malignant prostate epithelial tissue \textit{in vivo}, or \textit{in vitro} in primary cell cultures or cell lines, various \textsuperscript{13}C labeled substrates can be used. Although in principle we could model the metabolic pathways for any of these, for the purpose described above, we specifically focused on the application of {[1,6-\textsuperscript{13}C\textsubscript{2}]}glucose or {[3-\textsuperscript{13}C]}pyruvate (both substrates result in the same citrate labeling pattern, as shown below) and on {[2,5-\textsuperscript{13}C\textsubscript{2}]}glucose or {[2-\textsuperscript{13}C]}pyruvate (also both resulting in the same citrate labeling pattern). In practice, these substrates can be administered long enough to the cells (e.g. 48h) to reach a steady state of \textsuperscript{13}C labeling of citrate secreted in the extracellular fluid or incubation medium. Extracellular material can then be collected and analyzed by high resolution \textsuperscript{13}C (and \textsuperscript{1}H) NMR spectroscopy. Citrate resonances can be identified and quantified (e.g. using \textit{j}MRUI AMARES\supercite{Vanhamme1997,Naressi2001,Stefan2009}) for further use, e.g as input in the model. Using {[3,4-\textsuperscript{13}C\textsubscript{2}]}glucose or {[1-\textsuperscript{13}C]}pyruvate would not result in labeling of citrate via the PDC pathway, but only via the PC pathway.
\begin{figure}[!htb]
    \centering
    \includegraphics[width=0.6\textwidth]{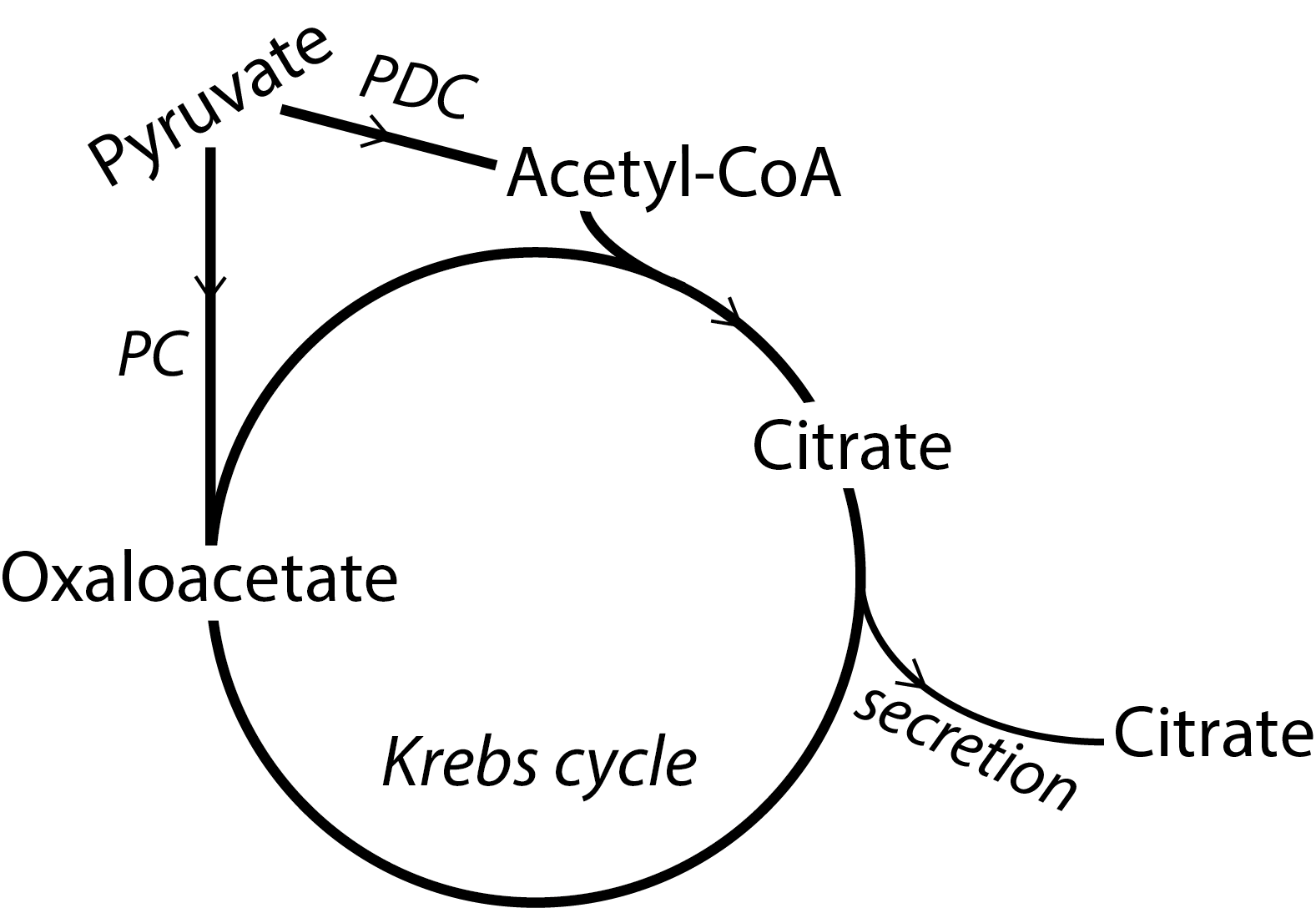}
    \caption{Schematic overview of the basic elements of the quantitative model for citrate secretion. Pyruvate can contribute carbons to the Krebs cycle either as Acetyl-CoA, through pyruvate dehydrogenase complex route (PDC), or in oxaloacetate, via the pyruvate carboxylase route (PC). Oxaloacetate and acetyl-CoA undergo a condensation reaction providing citrate, which is partially secreted. Note that this figure shows only the basics of our model and that other contributions to these pathways (anaplerotic and cataplerotic) are considered (see text).}
    \label{fig:14}
\end{figure}
\section{Quantitative model}
The distribution of \textsuperscript{13}C carbons in citrate, as determined from NMR spectra, reflects on the ratio of pyruvate going into the Krebs cycle via pyruvate carboxylase (as oxaloacetate) and pyruvate dehydrogenase (as acetyl-CoA) on one hand and on the fraction of molecules leaving the Krebs cycle, in every turn of this cycle, to be secreted into the lumen or incubation medium on the other hand (Figure \ref{fig:14}). Here we derive a quantitative model describing prostatic citrate production depending on a (constant) secretion fraction $c$ and on the fractions of carbons (either \textsuperscript{13}C labeled or unlabeled) entering via acetyl-CoA ($f^{PDC}$) or oxaloacetate ($f^{PC}$), with $f^{PC}+f^{PDC} =1$. The \textsuperscript{13}C labeling at the six different positions in citrate secreted and carbon dioxide produced at cycle turn $n$ is described by $\overrightarrow{C}(n)$.
    \begin{linenomath}\begin{equation}\overrightarrow{C}(n) = \begin{pmatrix}C_1(n)\\[5pt]C_2(n)\\[5pt]C_3(n)\\[5pt]C_4(n)\\[5pt]C_5(n)\\[5pt]C_6(n)\\[5pt]CO_2(n)\\[5pt]\end{pmatrix}c(1-c)^n\end{equation}\end{linenomath}
Since every Krebs cycle turn a fraction $c$ of citrate leaves the cycle and fraction $(1-c)$ remains in the cycle a factor $c(1-c)^n$ is included. For example, if at $n=3$ citrate molecules are secreted with 20\% label on $C_2$, 10\% on $C_4$ and 5\% on $C_5$ and $C_6$, and 3\% label is lost in carbon dioxide, this would be described by: \begin{linenomath}\begin{equation}\overrightarrow{C}(n) = \begin{pmatrix}0\\[5pt]0.2\\[5pt]0\\[5pt]0.1\\[5pt]0.05\\[5pt]0.05\\[5pt]0.03\\[5pt]\end{pmatrix}c(1-c)^3\end{equation}\end{linenomath}
Equations describing the labeling of each of the six carbons in secreted citrate are derived further on in this paper.

\subsection{Fate of \textsuperscript{13}C labels in the Krebs cycle}\label{sec:3}
\begin{figure}[!htb]
    \centering
    \includegraphics[width=1\textwidth]{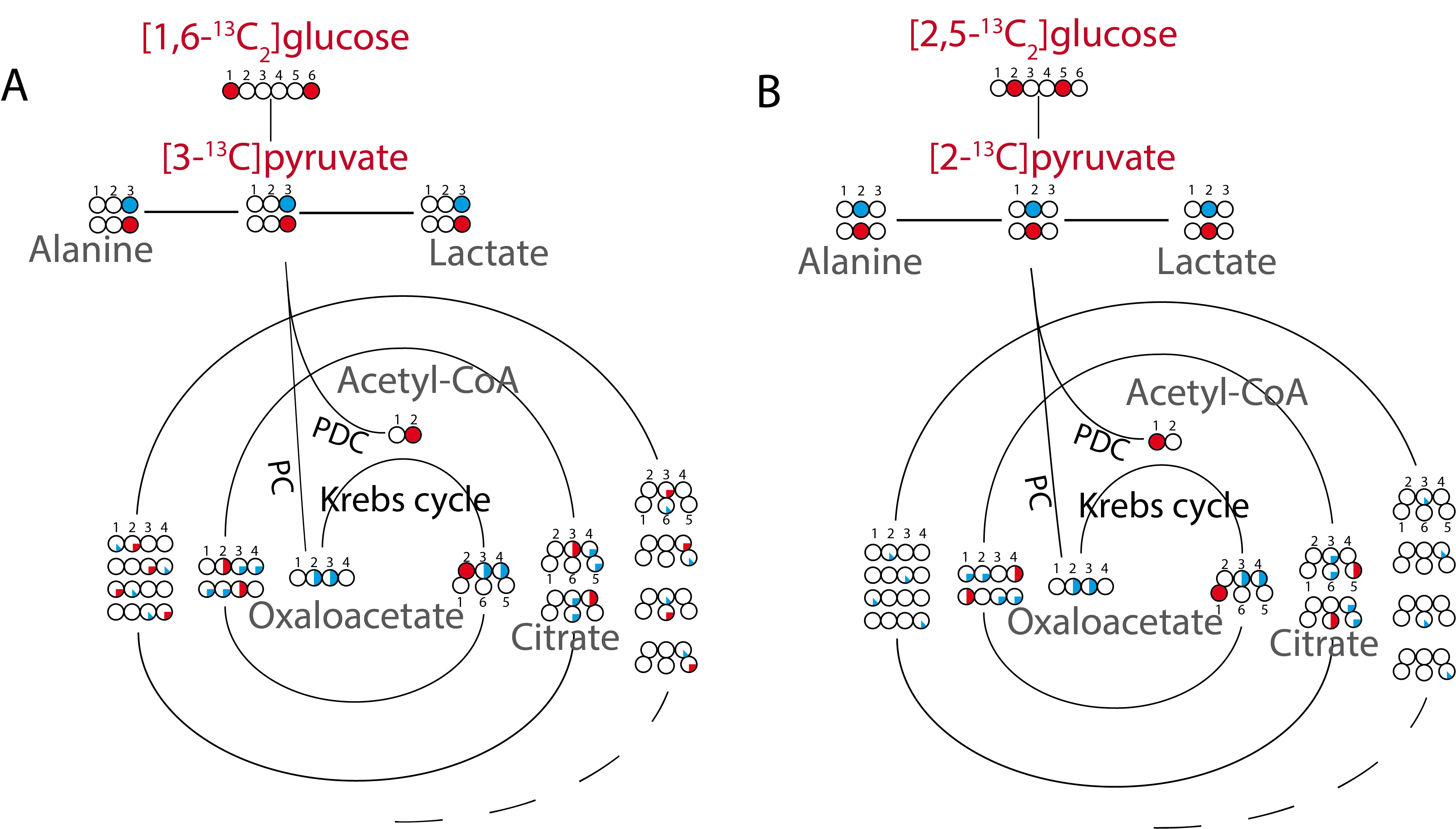}
    \caption{Schematic representation of the fate of \textsuperscript{13}C carbons entering the Krebs cycle during subsequent cycle turns. A.) The distribution of \textsuperscript{13}C labels originating from {[1,6-\textsuperscript{13}C\textsubscript{2}]}glucose or {[3-\textsuperscript{13}C]}pyruvate. (B) Idem for {[2,5-\textsuperscript{13}C\textsubscript{2}]}glucose or {[2-\textsuperscript{13}C]}pyruvate. Pyruvate carbons converted to oxaloacetate by pyruvate carboxylase (PC) are indicated by blue circles and those converted into acetyl-CoA by pyruvate dehydrogenase complex (PDC) by red circles. Rapid exchange between oxaloacetate and fumarate equally distributes the labeling of oxaloacetate via PC over the two middle carbons. In this figure the carbon numbers in each compound are indicated and for simplicity mitochondrial and cellular export of citrate is omitted. In this figure no new \textsuperscript{13}C label is flowing into the Krebs cycle during the subsequent cycle turns, this however is taken into account in the sections below.}
    \label{fig:13}
\end{figure}

When a \textsuperscript{13}C label enters the Krebs cycle it can end up at different positions in citrate (see Figure \ref{fig:13}). If a label continues along the Krebs cycle it can eventually end up in citrate again, but now at a different position, or even distributed equally over two citrate carbon positions. This is because succinate and fumarate are symmetrical intermediates in the Krebs cycle resulting in the indistinguishable labeling of C1 and C4 or of C2 and C3 in succinate and fumarate. Eventually this leads to an equal distribution of \textsuperscript{13}C labels at two carbon positions in citrate. The position of \textsuperscript{13}C labels at citrate carbons C1-C6 and oxaloacetate (OAA) at cycle turn $n+1$ versus $n$ are shown in Table \ref{tab:1}.
\begin{table}[!htb]
\begin{center}
$\begin{array}{ccccc}
    Citr(n)&\Rightarrow&OAA(n)&\Rightarrow&Citr(n+1) \\[5pt]\hline\\ [-1.5ex]
    C1&\Rightarrow&\frac{1}{2}C1+\frac{1}{2}C4&\Rightarrow&\frac{1}{2}C5+\frac{1}{2}C6\\[5pt]
    C2&\Rightarrow&\frac{1}{2}C2+\frac{1}{2}C3&\Rightarrow&\frac{1}{2}C3+\frac{1}{2}C4\\[5pt]
    C3&\Rightarrow&\frac{1}{2}C2+\frac{1}{2}C3&\Rightarrow&\frac{1}{2}C3+\frac{1}{2}C4\\[5pt]
    C4&\Rightarrow&\frac{1}{2}C1+\frac{1}{2}C4&\Rightarrow&\frac{1}{2}C5+\frac{1}{2}C6\\[5pt]
    C5&\Rightarrow&CO\textsubscript{2}\\[5pt]
    C6&\Rightarrow&CO\textsubscript{2}\\
\end{array}$
\caption{Shifting of carbons of citrate during the Krebs cycle. Shown are citrate (Citr) at Krebs cycle turn $n$, oxaloacetate (OAA) at cycle turn $n$ and citrate at cycle turn $n+1$. Cycle turn $n$ starts at the secretion of citrate and ends upon formation of new citrate in the subsequent cycle turn.}\label{tab:1}
\end{center}
\end{table}
\subsection{First labeling of citrate ($n=0$)} \label{sec:2}
\begin{figure}[!htb]
\centering
\small{
\begin{tikzpicture}[grow=right, parent anchor=east, child anchor=west, sloped, level distance=3cm, sibling distance=1cm]
    \node{{[2,5-\textsuperscript{13}C\textsubscript{2}]}Glc}
        child{
            node {{[2-\textsuperscript{13}C]}Pyr}        
            child {
                node(B){OxaAc}
                edge from parent[draw=none]
            }
            child{
                node(G) {{[1-\textsuperscript{13}C]}Ac-CoA}
                child[grow=-5.71,level distance=5cm] {
                    node(A) {{{[1-\textsuperscript{13}C]}Citr}}
                    edge from parent[draw=none]
                }
                child[grow=5.71,level distance=5cm] {
                    node(E) {{{[1,3-\textsuperscript{13}C\textsubscript{2}]}/{[1,4-\textsuperscript{13}C\textsubscript{2}]}Citr}}
                    edge from parent[draw=none]
                }
                edge from parent[->]
                node[below] {\small{\textit{PDC}}}
            }
            child {
                node(F) {{[2-\textsuperscript{13}C]}/{[3-\textsuperscript{13}C]}OxAc}
                child[grow=5.71,level distance=5cm] {
                node(C) {{{[3-\textsuperscript{13}C]}/{[4-\textsuperscript{13}C]}Citr}}
                    edge from parent[draw=none]
                }
                edge from parent[->]
                node[above] {\small{\textit{PC}}}
            }
            child {
                node(D) {Ac-CoA}
                edge from parent[draw=none]
            }
            edge from parent[->]
            node[above] {\small{\textit{Glycolysis}}}
        };
        \draw[<-] ([yshift=-0.1cm]A.west)  to [out=170,in=0] (B.east);
        \draw[<-] ([yshift=0.1cm]C.west)  to [out=-170,in=0] (D.east);
        \draw[<-] ([yshift=0.1cm]E.west) -- ([yshift=-0.1cm]F.east);
        \draw[<-] ([yshift=0.1cm]A.west) -- ([yshift=-0.1cm]G.east);
        \draw[<-] ([yshift=-0.1cm]E.west) -- ([yshift=0.1cm]G.east);
        \draw[<-] ([yshift=-0.1cm]C.west) -- ([yshift=0.1cm]F.east);
        \draw (8.2cm,0) node[rectangle, minimum height=3cm,minimum width=2cm, fill=gray,fill opacity=0.6] {
            \begin{tabular}{cc}
                Citrate\\
                synthase\\
            \end{tabular}};
\end{tikzpicture}}
\caption{For $n=0$ the \textsuperscript{13}C labels originating from {[2,5-\textsuperscript{13}C\textsubscript{2}]}glucose or {[2-\textsuperscript{13}C]}pyruvate can either end up in oxaloacetate (at C2 or C3) via pyruvate carboxylase or in acetyl-CoA (at C1) via pyruvate dehydrogenase complex before it ends up in citrate after the condensation reaction of acetyl-CoA and oxaloacetate. This can result in unlabeled (not shown), single labeled and double labeled citrate. Glc = glucose, Pyr = pyruvate, Ac-CoA = acetyl-CoA, OxAc = oxaloacetate.}
\label{fig:11}
\end{figure}
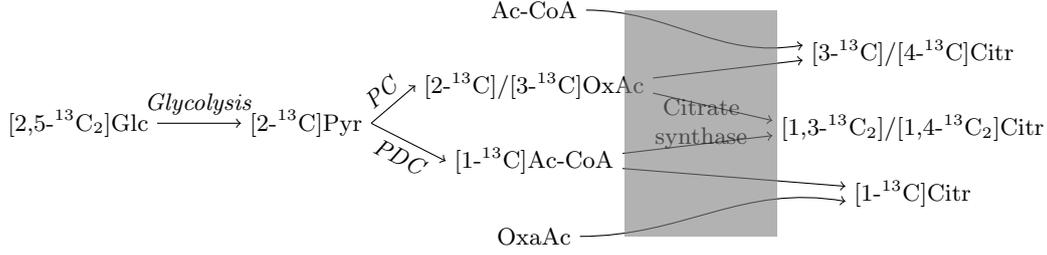

The (\textsuperscript{13}C) carbons of pyruvate can enter the Krebs cycle via the pyruvate carboxylase (PC) and the pyruvate dehydrogenase complex (PDC) route. We define Krebs cycle turn $n$ as the number of full Krebs cycle turns completed before secretion of citrate from the Krebs cycle. This means that if citrate is secreted right after it is labeled, this is Krebs cycle turn $n=0$. This is the only cycle turn where \textsuperscript{13}C carbons enter the Krebs cycle via PC because the oxaloacetate during the next Krebs cycle turns ($n>0$) originates from the citrate in the previous Krebs cycle turn that is not secreted. On the other hand, this oxaloacetate undergoes a condensation reaction with a \textsuperscript{13}C labeled acetyl-CoA forming citrate every Krebs cycle turn, so carbons enter the Krebs cycle via PDC every cycle turn. The first labeled citrate (at $n=0$) can have a different label depending on the labeling of pyruvate (or acetyl-CoA). For {[1,6-\textsuperscript{13}C\textsubscript{2}]}glucose or {[3-\textsuperscript{13}C]}pyruvate the route of labeling of citrate is shown in Figures \ref{fig:13} and \ref{fig:10}. The \textsuperscript{13}C labeled oxaloacetate (OAA) and \textsuperscript{13}C labeled acetyl-CoA can, together with unlabeled oxaloacetate or acetyl-CoA, undergo a condensation reaction to produce unlabeled, single- or double \textsuperscript{13}C labeled citrate. This results in citrate labeled at C2 (PDC) and/or C1/4 (PC) respectively at $n=0$. The labeling of oxaloacetate is distributed over two carbons due to rapid exchange between oxaloacetate, malate and (symmetrical) fumarate.\supercite{Magnusson1991,Merritt2011} If {[2,5-\textsuperscript{13}C\textsubscript{2}]}glucose or {[2-\textsuperscript{13}C]}pyruvate are applied as substrate the labeling paths as shown in Figure \ref{fig:13}B (or Figure \ref{fig:11}) are followed, resulting in citrate labeling at C1 (PDC) and/or C2/3 (PC) respectively at $n=0$.
\bigskip\\
Note here that if $P$ and $A$ were to be the fractions of \textsuperscript{13}C label ending up in oxaloacetate via PC or in acetyl-CoA via PDC respectively, the resulting distribution of \textsuperscript{13}C label over the three citrate positions that are possibly labeled at $n=0$ is equal to $\frac{P}{2}$ at two positions and $A$ at the other. For the {[1,6-\textsuperscript{13}C\textsubscript{2}]}glucose or {[3-\textsuperscript{13}C]}pyruvate experiment this would be $A$ label at C2 and $P$ labels distributed over C3 and C4. For the \textsuperscript{13}C {[2,5-\textsuperscript{13}C\textsubscript{2}]}glucose or {[2-\textsuperscript{13}C]}pyruvate experiment this results in $A$ label at the C1 position and $P$ label distributed over the C3 and C4 position.
\begin{figure}[!htb]
\centering
\small{
\begin{tikzpicture}[grow=right, parent anchor=east, child anchor=west, sloped, level distance=3cm, sibling distance=1cm]
    \node{{[1,6-\textsuperscript{13}C\textsubscript{2}]}Glc}
        child{
            node {{[3-\textsuperscript{13}C]}Pyr}        
            child {
                node(B){OxaAc}
                edge from parent[draw=none]
            }
            child{
                node(G) {{[2-\textsuperscript{13}C]}Ac-CoA}
                child[grow=-5.71,level distance=5cm] {
                    node(A) {{{[2-\textsuperscript{13}C]}Citr}}
                    edge from parent[draw=none]
                }
                child[grow=5.71,level distance=5cm] {
                    node(E) {{{[2,4-\textsuperscript{13}C\textsubscript{2}]}/{[2,3-\textsuperscript{13}C\textsubscript{2}]}Citr}}
                    edge from parent[draw=none]
                }
                edge from parent[->]
                node[below] {\small{\textit{PDC}}}
            }
            child {
                node(F) {{[3-\textsuperscript{13}C]}/{[2-\textsuperscript{13}C]}OxAc}
                child[grow=5.71,level distance=5cm] {
                node(C) {{{[4-\textsuperscript{13}C]}/{[3-\textsuperscript{13}C]}Citr}}
                    edge from parent[draw=none]
                }
                edge from parent[->]
                node[above] {\small{\textit{PC}}}
            }
            child {
                node(D) {Ac-CoA}
                edge from parent[draw=none]
            }
            edge from parent[->]
            node[above] {\small{\textit{Glycolysis}}}
        };
        \draw[<-] ([yshift=-0.1cm]A.west)  to [out=170,in=0] (B.east);
        \draw[<-] ([yshift=0.1cm]C.west)  to [out=-170,in=0] (D.east);
        \draw[<-] ([yshift=0.1cm]E.west) -- ([yshift=-0.1cm]F.east);
        \draw[<-] ([yshift=0.1cm]A.west) -- ([yshift=-0.1cm]G.east);
        \draw[<-] ([yshift=-0.1cm]E.west) -- ([yshift=0.1cm]G.east);
        \draw[<-] ([yshift=-0.1cm]C.west) -- ([yshift=0.1cm]F.east);
        \draw (8.2cm,0) node[rectangle, minimum height=3cm,minimum width=2cm, fill=gray,fill opacity=0.6] {
            \begin{tabular}{cc}
                Citrate\\
                synthase\\
            \end{tabular}};
\end{tikzpicture}}
\caption{For $n=0$ the \textsuperscript{13}C labels originating from {[1,6-\textsuperscript{13}C\textsubscript{2}]}glucose or {[3-\textsuperscript{13}C]}pyruvate end up in oxaloacetate (at C2 or C3) via pyruvate carboxylase or in acetyl-CoA (at C2) via pyruvate dehydrogenase complex before ending up in citrate after the condensation reaction of acetyl-CoA and oxaloacetate. This can result in unlabeled (not shown), single labeled and double labeled citrate. Glc = glucose, Pyr = pyruvate, Ac-CoA = acetyl-CoA, OxAc = oxaloacetate.}
\label{fig:10}
\end{figure}
\subsection{Pyruvate carboxylase versus pyruvate dehydrogenase complex}
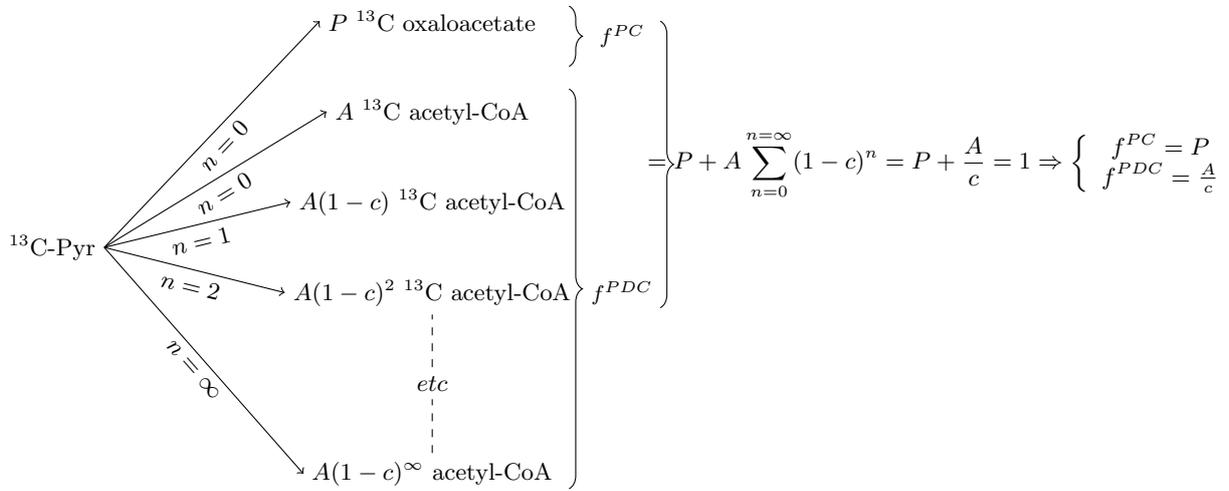
\begin{figure}[!htb]
\centering
\small{
\makebox[\textwidth]{
\begin{tikzpicture}[parent anchor=east, child anchor=west, grow=right, sloped, level distance=5cm, sibling distance=1.2cm]
    \node{\textsuperscript{13}C-Pyr}
        child{
            node (b) {$A(1-c)^\infty$ acetyl-CoA}
            edge from parent[->]
            node[below] {\small{\textit{$n=\infty$}}}
        }
        child[ultra thin, dotted]{
            node[thin] (d) {$etc$}
            edge from parent[draw=none,anchor=south]
        }
        child{
            node (c) {$A(1-c)^2$ \textsuperscript{13}C acetyl-CoA} 
            edge from parent[->]
            node[below] {\small{\textit{$n=2$}}}
        }
        child{
            node {$A(1-c)$ \textsuperscript{13}C acetyl-CoA}        
            edge from parent[->]
            node[below] {\small{\textit{$n=1$}}}
        }
        child{
            node {$A$ \textsuperscript{13}C acetyl-CoA}        
            edge from parent[->]
            node[below] {\small{\textit{$n=0$}}}
        }
        child{
            node (a) {$P$ \textsuperscript{13}C oxaloacetate} 
            edge from parent[->]
            node[below] {\small{\textit{$n=0$}}}
        };
        \draw [decorate,decoration={brace,amplitude=5pt,mirror,raise=0cm},yshift=0cm] (6.8cm,-3.2cm) -- (6.8cm,2.1 cm);
        \draw [decorate,decoration={brace,amplitude=5pt,mirror,raise=0cm},yshift=0cm] (6.8cm,2.4cm) -- (6.8cm,3.2cm);
        \draw node(k)[xshift=7.5cm,yshift=-0.6cm] {\footnotesize $f^{PDC}$};
        \draw node(l)[xshift=7.5cm,yshift=2.8cm] {\footnotesize $f^{PC}$};
        \draw[dashed] (b) -- (d);
        \draw[dashed] (d) -- (c);
        \draw [decorate,decoration={brace,amplitude=5pt,mirror,raise=0cm},xshift=0cm] (8cm,-.8cm) -- (8cm,3cm);
        \draw node(l)[xshift=11.7cm,yshift=1.1cm]  {$\displaystyle =P+A\sum\limits_{n=0}^{n=\infty}(1-c)^n = P+\frac{A}{c}=1\displaystyle \Rightarrow \left\{\begin{array}{cc} f^{PC}=P\\ f^{PDC}=\frac{A}{c}\end{array}\right.$};
\end{tikzpicture}}}
\caption{Fate of \textsuperscript{13}C labels of pyruvate entering the Krebs cycle and ending up in citrate. Fraction $P=f^{PC}$ enters via the pyruvate carboxylase pathway as oxaloacetate (only at $n=0$), fraction $f^{PDC}=\frac{A}{c}$ flows into the Krebs cycle via the pyruvate dehydrogenase complex pathway as acetyl-CoA.}\label{fig:9}
\end{figure}

\begin{figure}[!ht]
    \centering
    \includegraphics[width=1\textwidth]{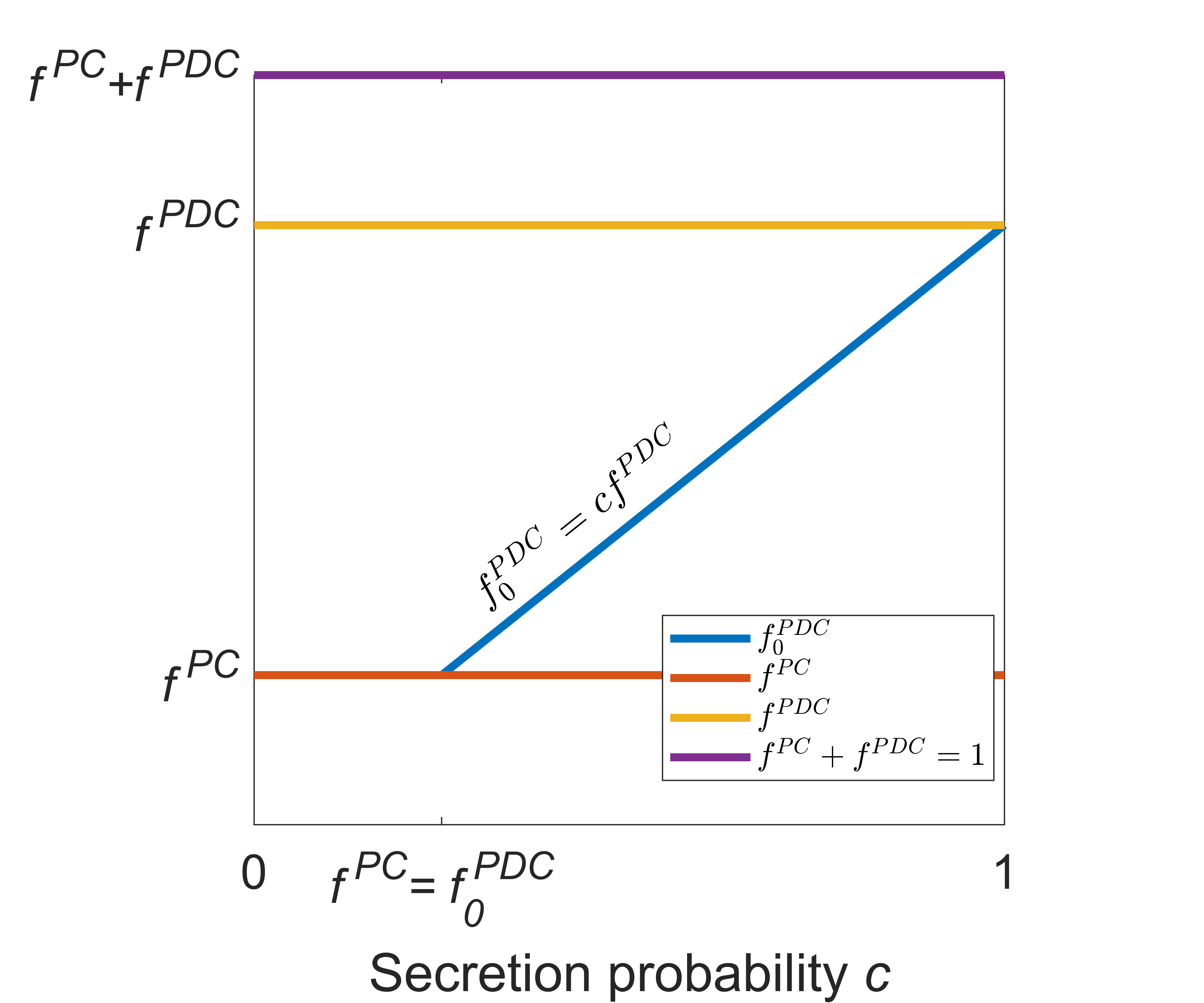}
\caption{Fractions of carbons diverging via pyruvate carboxylase $f^{PC}$ and via pyruvate dehydrogenase complex $f^{PDC}$ versus secretion fraction $c$.
\newline The sum of $f^{PC}$ and $f^{PDC}$ is equal to $1$. The initial fraction (at $n=0$) of carbons entering the Krebs cycle via PDC ($f^{PDC}_0$), depends on the value of $c$, since $f^{PDC}=f^{PDC}_0\sum\limits_{n=0}^{n=\infty}(1-c)^n=\frac{f^{PDC}_0}{c}$. Next to this, $f^{PDC}_0 \leq f^{PC}$, since for every oxaloacetate labeled via the PC route, there is an acetyl-CoA needed (labeled by PDC) at $n=0$, making the blue line start at $f^{PC}=cf^{PDC}$\label{fig:1}}
\end{figure}

To determine the fractional contributions of the PC and PDC routes to the final citrate pool in the luminal fluid or incubation medium it is essential to realize that depending on how many Krebs cycle turns a \textsuperscript{13}C label completes before being secreted, the contribution is different. Let's start with fraction A \textsuperscript{13}C labels getting into citrate via PDC and fraction P \textsuperscript{13}C labels getting into citrate via PC. Subsequently fraction $c$ of the citrate will be secreted (we have defined this as Krebs cycle turn $n=0$) and fraction $(1-c)$ continues in the Krebs cycle. Now $(1-c)$ carbon skeletons will be labeled again upon condensation of the carbon skeleton (oxaloacetate) with $(1-c)$ \textsuperscript{13}C acetyl-CoA forming $(1-c)$ citrate. Of this citrate again fraction $c$ is secreted (at cycle turn $n=1$) and fraction $(1-c)$ continues down the Krebs cycle and the process repeats itself. This is summarized in Figure \ref{fig:9}. Pyruvate carboxylase only contributes to the \textsuperscript{13}C labeling of the citrate pool in the luminal fluid or incubation medium at $n=0$ and pyruvate dehydrogenase complex contributes to the citrate pool during all Krebs cycle turns. In general, during cycle turn $n>0$ there will be $A(1-c)^n$ labels entering the cycle and $Ac(1-c)^n$ citrate molecules will be secreted. Only at $n=0$ there will be $A+P$ labels entering the cycle and $c(A+P)$ citrate molecules will be secreted. The sum of all fractions of \textsuperscript{13}C labels entering the Krebs cycle has to be equal to $f^{PC}+f^{PDC} =1$ and since the only time labels originating from the PC route can enter the cycle is at turn $n=0$, we can easily see that the fraction $f^{PC}$ of pyruvate carbons is equal to the fraction $P$ of \textsuperscript{13}C pyruvate carbons described above. As fraction $f^{PDC}$ has to be equal to the sum of fractions of \textsuperscript{13}C labels entering via acetyl-CoA, this comes down to the following sum:
\begin{linenomath}\begin{equation}
    f^{PDC} = A\sum\limits_{n=0}^{n=\infty}(1-c)^n = \frac{A}{c}
\end{equation}\end{linenomath}
This means that if at $n=0$ the fraction $f^{PC}$ of pyruvate \textsuperscript{13}C labels enter the cycle via the PC route and fraction $A=cf^{PDC}$ enters the cycle via the PDC route, summed over all possible Krebs cycle turns a total of $f^{PC} + \frac{cf^{PDC}}{c} = f^{PC} + f^{PDC} = 1$ \textsuperscript{13}C of the labels enter the Krebs cycle (Figure \ref{fig:1}).\\

\subsection{Average number of completed Krebs cycle turns before citrate secretion} \label{sec:7}
The average number of Krebs cycle turns before secretion of citrate (This is not the same as the average number of Krebs cycle turns a \textsuperscript{13}C carbon completes, see Appendices for the average number of cycle turns a \textsuperscript{13}C carbon completes) depends on the secretion fraction $c$. To calculate this average number of cycle turns $<n>$ the sum can be calculated of the fractions secreted every cycle $c(1-c)^n$ multiplied with their corresponding Krebs cycle turn indices $n$.
\begin{linenomath}\begin{equation}
    <n>=c\sum\limits_{n=0}^{n=\infty}n(1-c)^n=\frac{1-c}{c}\label{eq:1}
\end{equation}\end{linenomath}
The average number of cycle turns is equal to the ratio of the fraction of molecules that stay in the Krebs cycle $(1-c)$ over the fraction of molecules that are secreted $c$.
\subsection{Effect of \textsuperscript{13}C enrichment} \label{sec:4}
Until now, we assumed that all pyruvate or glucose carbons are \textsuperscript{13}C labeled and that both the pyruvate pool and acetyl-CoA pool are fully labeled. This probably doesn't reflect the true situation in cells or tissue because of influx of unlabeled pyruvate (e.g. via Malic Enzyme or PEPCK) or acetyl-CoA (e.g. via beta oxidation). To take this into account, introducing enrichment factors in the equations is needed. Suppose only fraction $\epsilon_p$ of pyruvate and fraction $\epsilon_a$ of acetyl-CoA is \textsuperscript{13}C labeled, then fractions $\epsilon_p\epsilon_a cf^{PDC}$ (via pyruvate dehydrogenase complex) and $\epsilon_p f^{PC}$ (via pyruvate carboxylase) of \textsuperscript{13}C carbons will enter the Krebs cycle at $n=0$, $\epsilon_p\epsilon_a cf^{PDC}(1-c)$ at $n=1$, $\epsilon_p\epsilon_a cf^{PDC}(1-c)^2$ at $n=2$ and so forth. This means that the contributions of the individual cycle turns to the sum (final \textsuperscript{13}C distribution over citrate carbons) are all multiplied by $\epsilon_p$, but only the carbons going via the PDC route (involving acetyl-CoA) are also multiplied with $\epsilon_a$. This only holds if the values of $\epsilon_p$ and $\epsilon_a$ are not time-dependent. If high enough concentrations or constant inflow of \textsuperscript{13}C substrate is provided, we can safely assume this to be true during most of the infusion or incubation period. For sake of simplicity we will first derive equations for the \textsuperscript{13}C distribution of citrate assuming both enrichment fractions to be 1, and later reintroduce these fractions.
\subsection{Assumptions made in deriving the quantitative model} \label{sec:10}
In deriving this model we assumed firstly that \textsuperscript{13}C carbons of pyruvate can only enter the Krebs cycle via the pyruvate carboxylase (PC) route, resulting in labeled oxaloacetate or via the pyruvate dehydrogenase complex (PDC) route, resulting in labeled acetyl-CoA. Other pathways are assumed to be negligble. Secondly we assume that a dominant part of molecules leaving the Krebs cycle is due to citrate secretion, and therefore we later define the total fraction of citrate and other molecules (citrate equivalents) to be an apparent citrate secretion fraction $d$.  If there are no other cataplerotic pathways involved, this fraction $d$ is equal to the true citrate secretion fraction $c$. This is assumed during the derivation of the equations below, but fraction $d$ is reintroduced afterwards for completeness. Thirdly, we assume that the \textsuperscript{13}C-enrichments of both the pyruvate and acetyl-CoA pool are constant over time. Deriving the equations below we assume both to be 1, but take them into account later. We also assume that \textsuperscript{13}C-labeled citrate or citrate equivalents (Krebs cycle metabolites) diverged from the Krebs cycle only to reenter the Krebs cycle through exchange, and not at a different position in the Krebs cycle after being converted into another Krebs cycle metabolite, so no short-cuts within the Krebs cycle. The effect of these short-cuts is assumed to be negligible and therefore ignored in this manuscript. We also assume that the fraction of carbon skeletons diverging from the Krebs cycle every cycle turn remains constant over the time period of the experiments.If labeled citrate is used for lipid synthesis, this can result in backflow of labels into the Krebs cycle via oxaloacetate, this effect is expected to be negligible and is therefore not taken into account.\\
Important to note is: Anaplerotic inflow of (unlabeled) carbons from e.g. glutamine, aspartate or malate during supplementation with {[1,6-\textsuperscript{13}C\textsubscript{2}]}glucose or {[2-\textsuperscript{13}C]}pyruvate do not contribute to the \textsuperscript{13}C-labeling pattern of citrate since they do not contribute (see assumption below) \textsuperscript{13}C-carbons to either the \textsuperscript{13}C pools of pyruvate or acetyl-CoA. Next to this, exchange of Krebs cycle metabolites with metabolites outside the Krebs cycle does not change the labeling pattern of the citrate that is secreted, but do lead to loss of citrate equivalents. For example, exchange of oxaloacetate with a potential aspartate pool or the exchange of $\alpha$-ketoglutarate with the glutamate pool does not change the labeling pattern of the $\alpha$-ketoglutarate that continues in the Krebs cycle. Even if it continues to be converted to glutamine and back. It does however lead to loss of carbon skeletons (citrate equivalents) if there's net flux of $\alpha$-ketoglutarate to the glutamate pool. These cataplerotic pathways are thus contributing to the apparent citrate secretion fraction $d$ and this apparent citrate secretion fraction is thus an overestimation of the real citrate secretion fraction. Fast exchange between malate, fumarate and oxaloacetate results in full scrambling of the carbons in oxaloacetate produced via the pyruvate carboxylase pathway. And the fast exchange between citrate, isocitrate and $\alpha$-ketoglutarate results in loss of citrate C6 labeling, taken into account by introducing a dilution factor $\epsilon_c$ later on.

%% file: p03_Derivation_glc_16.tex
\subsection{Providing {[1,6-\textsuperscript{13}C\textsubscript{2}]}glucose or {[3-\textsuperscript{13}C]}pyruvate to tissue/cells} \label{sec:5}
\begin{figure}[!htb]
    \centering
    \includegraphics[width=1\textwidth]{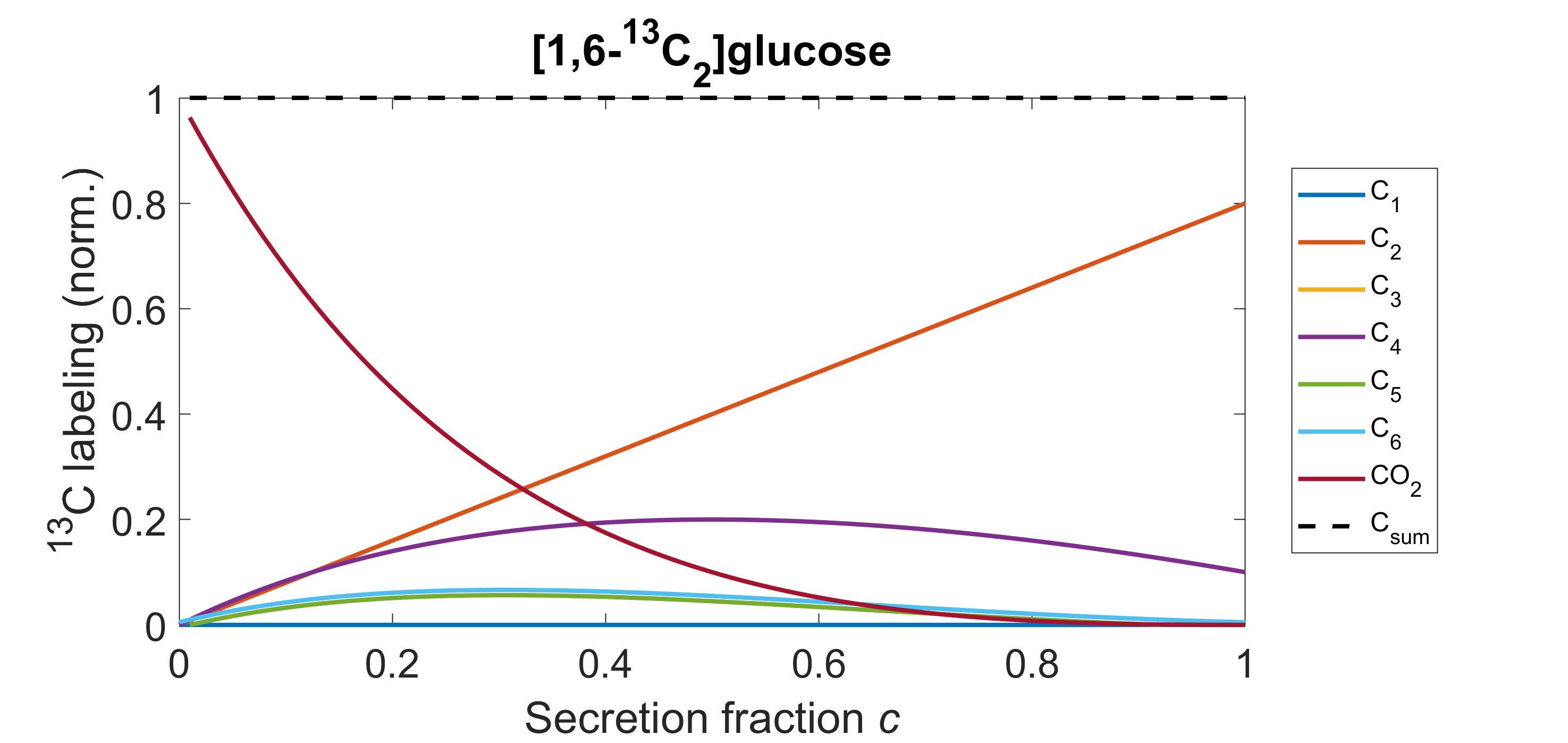}
    \caption{\textsuperscript{13}C distribution over the six citrate carbons versus secretion fraction $c$ after providing {[1,6-\textsuperscript{13}C\textsubscript{2}]}glucose or {[3-\textsuperscript{13}C]}pyruvate to prostate epithelial tissue or cells. In this graph the fraction of carbons following the pyruvate carboxylase pathway, $f^{PC}$, is assumed to be  $f^{PC}=0.2$. The \textsuperscript{13}C labeling is normalized to the total amount of \textsuperscript{13}C label going into the Krebs cycle.}
    \label{fig:2}
\end{figure}

To analyze how glucose or pyruvate contribute to the production and secretion of citrate, {[1,6-\textsuperscript{13}C\textsubscript{2}]}glucose or {[3-\textsuperscript{13}C]}pyruvate can be provided to epithelial cells or tissue for a longer period of time. Glucose is taken up by the cells and converted into pyruvate during glycolysis. In case of {[1,6-\textsuperscript{13}C\textsubscript{2}]}glucose the \textsuperscript{13}C label will be located at the C3 position of pyruvate as shown in section \textit{"\nameref{sec:2}"}. Subsequently this label can end up at the oxaloacetate C2 and C3 position, or acetyl-CoA C2 position. The first will result in labeling at C3 or C4 of citrate, the latter at C2. So, the \textsuperscript{13}C carbon distribution in citrate at cycle turn $n=0$ will look like (neglecting carbon dioxide for now):
\begin{linenomath}\begin{equation}
    \overrightarrow{C}(n=0)=\begin{pmatrix}0\\cf^{PDC}\\\frac{1}{2}f^{PC}\\\frac{1}{2}f^{PC}\\0\\0\end{pmatrix}c    
\end{equation}\end{linenomath}
After one turn the distribution will have shifted according to Table \ref{tab:1} and a new label is entered at the C2 position via PDC.
\begin{linenomath}\begin{equation}
    \overrightarrow{C}(n=1)=\begin{pmatrix}0\\[5pt]cf^{PDC}\\[5pt]\frac{c}{2}f^{PDC}+\frac{1}{4}f^{PC}\\[5pt]\frac{c}{2}f^{PDC}+\frac{1}{4}f^{PC}\\[5pt]\frac{1}{4}f^{PC}\\[5pt]\frac{1}{4}f^{PC}\\[5pt]\end{pmatrix}c(1-c)
\end{equation}\end{linenomath}
We assume here that all the label entering at the C2 position is \textsuperscript{13}C enriched as discussed in section \textit{"\nameref{sec:4}"}. For cycle turns $n\geq2$ the distribution at cycle $n$ can be written in a more general way as:
\begin{linenomath}\begin{equation}
\overrightarrow{C}(n\geq2)=\begin{pmatrix}0\\[5pt]cf^{PDC}\\[5pt]
    \sum\limits_{i=1}^{i=n}\frac{(1-c)^n}{2^i}cf^{PDC}+\frac{(1-c)^n}{2^{n+1}}f^{PC}\\[5pt]
    \sum\limits_{i=1}^{i=n}\frac{(1-c)^n}{2^i}cf^{PDC}+\frac{(1-c)^n}{2^{n+1}}f^{PC}\\[5pt]
    \sum\limits_{i=2}^{i=n}\frac{(1-c)^n}{2^i}cf^{PDC}+\frac{(1-c)^n}{2^{n+1}}f^{PC}\\[5pt]
    \sum\limits_{i=2}^{i=n}\frac{(1-c)^n}{2^i}cf^{PDC}+\frac{(1-c)^n}{2^{n+1}}f^{PC}\\[5pt]\end{pmatrix}c
\end{equation}\end{linenomath}
We would like to know the distribution of \textsuperscript{13}C labels over the six citrate carbons in the extracellular fluid or incubation medium when the citrate \textsuperscript{13}C distribution is in a steady state. This distribution can be calculated by summing the contributions of citrate secreted during each individual Krebs cycle turn $n$ to the total secreted citrate:
\begin{linenomath}\begin{align}
\overrightarrow{^{13}C}_{distr}^{1,6\mbox{-}glc}=\sum\limits_{n=0}^{n=\infty}\overrightarrow{C}(n)=\begin{pmatrix}0\\[5pt]
    cf^{PDC}\sum\limits_{j=0}^{j=\infty}(1-c)^j\\[5pt]
    cf^{PDC}\sum\limits_{j=1}^{j=\infty}\sum\limits_{i=1}^{i=j}\frac{(1-c)^j}{2^i}+f^{PC}\sum\limits_{j=0}^{j=\infty}\frac{(1-c)^j}{2^{j+1}}\\[5pt]
    cf^{PDC}\sum\limits_{j=1}^{j=\infty}\sum\limits_{i=1}^{i=j}\frac{(1-c)^j}{2^i}+f^{PC}\sum\limits_{j=0}^{j=\infty}\frac{(1-c)^j}{2^{j+1}}\\[5pt]
    cf^{PDC}\sum\limits_{j=2}^{j=\infty}\sum\limits_{i=2}^{i=j}\frac{(1-c)^j}{2^i}+f^{PC}\sum\limits_{j=0}^{j=\infty}\frac{(1-c)^j}{2^{j+1}}\\[5pt]
    cf^{PDC}\sum\limits_{j=2}^{j=\infty}\sum\limits_{i=2}^{i=j}\frac{(1-c)^j}{2^i}+f^{PC}\sum\limits_{j=0}^{j=\infty}\frac{(1-c)^j}{2^{j+1}}\\[5pt]\end{pmatrix}c=\begin{pmatrix}0\\[5pt]f^{PDC}\\[5pt]
    \frac{1-c}{1+c}f^{PDC}+\frac{1}{1+c}f^{PC}\\[5pt]
    \frac{1-c}{1+c}f^{PDC}+\frac{1}{1+c}f^{PC}\\[5pt]
    \frac{1-c}{2(1+c)}\left((1-c)f^{PDC}+f^{PC}\right)\\[5pt]
    \frac{1-c}{2(1+c)}\left((1-c)f^{PDC}+f^{PC}\right)\\[5pt]\end{pmatrix}c
\end{align}\end{linenomath}
These equations give the \textsuperscript{13}C label distribution over the six carbons in secreted citrate. This distribution only depends oction $c$ and the fraction of carbon entering the Krebs cycle via the pyruvate carboxylase route, $f^{PC}$, since $f^{PDC}=1-f^{PC}$. The \textsuperscript{13}C distribution for $f^{PC}=0.2$ is plotted against $c$ in figure \ref{fig:2}, the distribution of \textsuperscript{13}C labels over the six citrate carbons (and the fraction ending up in CO\textsubscript{2}) is strongly dependent on secretion fraction $c$. The distribution over the different carbons serves as a fingerprint for secretion fraction $c$.

%% file: p04_Derivation_pyr_2.tex
\subsection{Providing {[2,5-\textsuperscript{13}C\textsubscript{2}]}glucose or {[2-\textsuperscript{13}C]}pyruvate to tissue/cells} \label{sec:9}
\begin{figure}[!htb]
    \centering
    \includegraphics[width=1\textwidth]{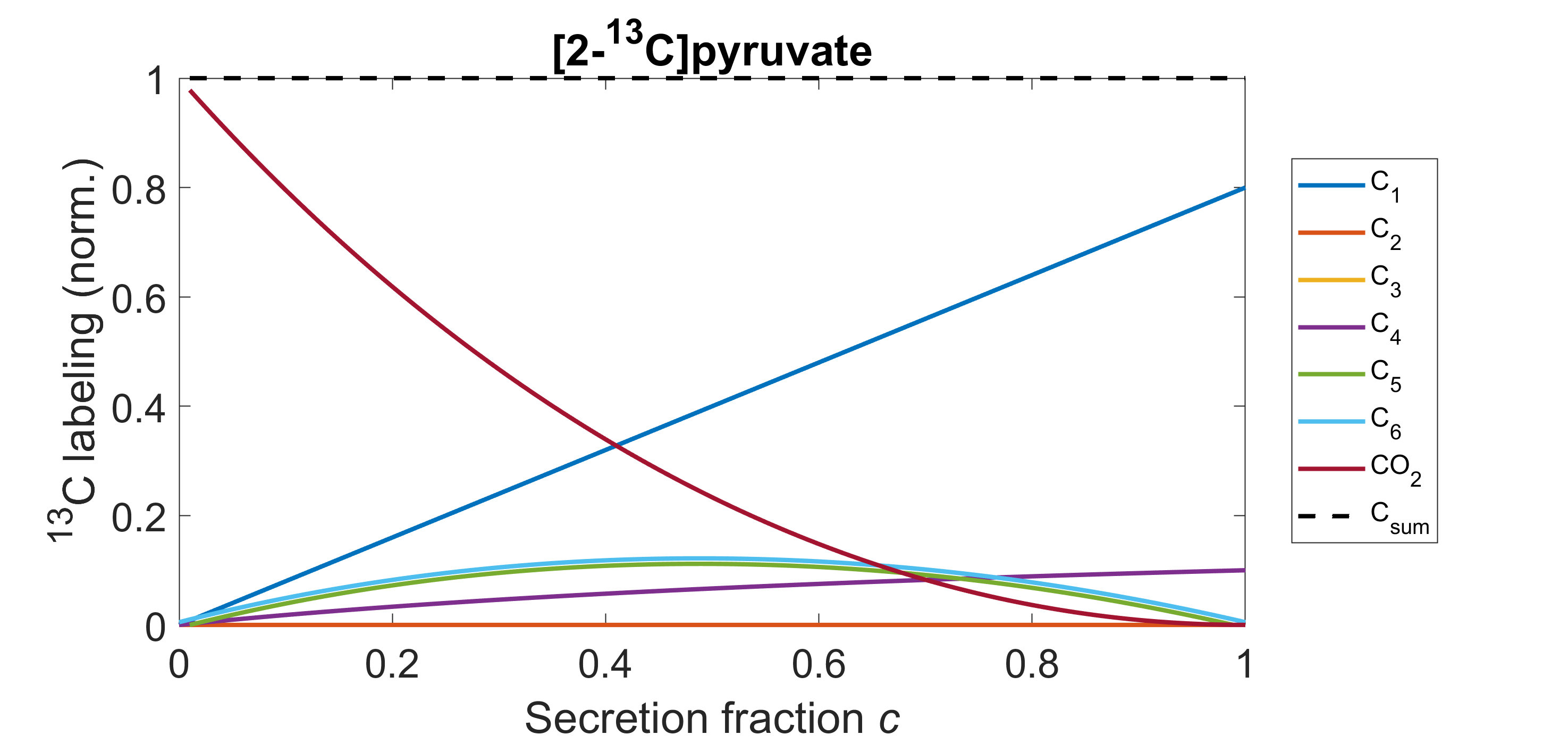}
    \caption{\textsuperscript{13}C distribution over the six citrate carbons versus secretion fraction $c$ after providing {[2,5-\textsuperscript{13}C\textsubscript{2}]}glucose or {[2-\textsuperscript{13}C]}pyruvate to prostate epithelial tissue or cells. In this graph the fraction of carbons following the pyruvate carboxylase pathway, $f^{PC}$, is assumed to be  $f^{PC}=0.2$. The \textsuperscript{13}C labeling is normalized to the total amount of \textsuperscript{13}C label going into the Krebs cycle. Enrichments $\epsilon_p$, $\epsilon_a$ and dilution fraction $\epsilon_c$ are assumed to be 1 here and no efflux of metabolites is taken into account apart from citrate secretion ($c=d$)}
    \label{fig:3}
\end{figure}

Another way to determine how differently labeled glucose or pyruvate contribute to the production of citrate, {[2,5-\textsuperscript{13}C\textsubscript{2}]}glucose or {[2-\textsuperscript{13}C]}pyruvate could be provided to epithelial tissue or prostate epithelial cells for a long enough period of time. Pyruvate is taken up by the cells and in this case the \textsuperscript{13}C label will be located at a different pyruvate carbon position compared to the experiments with {[1,6-\textsuperscript{13}C\textsubscript{2}]}glucose or {[3-\textsuperscript{13}C]}pyruvate (vide supra). When this \textsuperscript{13}C label enters the Krebs cycle, it can end up at the oxaloacetate C2 and C3 position, via the pyruvate carboxylase route, or at the acetyl-CoA C1 position via the pyruvate dehydrogenase complex route. The first will result in a first labeling at C3 and C4 of citrate, the latter at C1. So, the \textsuperscript{13}C carbon distribution in citrate at cycle turn $n=0$ will look like (again, neglecting carbon dioxide for now):
\begin{linenomath}\begin{equation}
    \overrightarrow{C}(n=0)=\begin{pmatrix}cf^{PDC}\\0\\\frac{1}{2}f^{PC}\\\frac{1}{2}f^{PC}\\0\\0\end{pmatrix}c    
\end{equation}\end{linenomath}
After one turn this distribution will have shifted again as described in section \textit{"\nameref{sec:5}"} (see also Table \ref{tab:1}) and new label is added, via the PDC route, at the C1 (via PDC) position of citrate every cycle turn.
\begin{linenomath}\begin{equation}
    \overrightarrow{C}(n=1)=\begin{pmatrix}cf^{PDC}\\[5pt]0\\[5pt]
    \frac{1}{4}f^{PC}\\[5pt]
    \frac{1}{4}f^{PC}\\[5pt]
    \frac{c}{2}f^{PDC}+\frac{1}{4}f^{PC}\\[5pt]
    \frac{c}{2}f^{PDC}+\frac{1}{4}f^{PC}\end{pmatrix}c(1-c)
\end{equation}\end{linenomath}
From cycle turn $n\geq2$ on the distribution at cycle turn $n$ can be written in a more general way as
\begin{linenomath}\begin{equation}
\overrightarrow{C}(n\geq2)=\begin{pmatrix}cf^{PDC}\\[5pt]0\\[5pt]
    \frac{1}{2^{n+1}}f^{PC}\\[5pt]
    \frac{1}{2^{n+1}}f^{PC}\\[5pt]
    \frac{1}{2^{n+1}}f^{PC}+\frac{c}{2}f^{PDC}\\[5pt]
    \frac{1}{2^{n+1}}f^{PC}+\frac{c}{2}f^{PDC}\end{pmatrix}c(1-c)^n
\end{equation}\end{linenomath}
The distribution of \textsuperscript{13}C carbons over the carbons in the secreted citrate can be calculated again by calculating the sum of the contributions of all individual Krebs cycle turns $n$
\begin{linenomath}\begin{equation}\begin{split}
\overrightarrow{^{13}C}_{distr}^{glc\mbox{-}2,5}=\sum\limits_{n=0}^{n=\infty}\overrightarrow{C}(n)=&\begin{pmatrix}c\sum\limits_{n=0}^{n=\infty}(1-c)^nf^{PDC}\\[5pt]0\\
    \frac{1}{2}\sum\limits_{i=0}^{i=\infty}\left(\frac{1-c}{2}\right)^if^{PC}\\[5pt]
    \frac{1}{2}\sum\limits_{i=0}^{i=\infty}\left(\frac{1-c}{2}\right)^if^{PC}\\[5pt]
    \frac{c}{2}\sum\limits_{i=1}^{i=\infty}(1-c)^if^{PDC}+\frac{1}{2}\sum\limits_{i=1}^{i=\infty}\left(\frac{1-c}{2}\right)^if^{PC}\\[5pt]
    \frac{c}{2}\sum\limits_{i=1}^{i=\infty}(1-c)^if^{PDC}+\frac{1}{2}\sum\limits_{i=1}^{i=\infty}\left(\frac{1-c}{2}\right)^if^{PC}\end{pmatrix}c\\
=&\quad\begin{pmatrix}f^{PDC}\\[5pt]0\\[5pt]
    \frac{1}{1+c}f^{PC}\\[5pt]
    \frac{1}{1+c}f^{PC}\\[5pt]
    \frac{1-c}{2}\left(\frac{1}{1+c}f^{PC}+f^{PDC}\right)\\[5pt]
    \frac{1-c}{2}\left(\frac{1}{1+c}f^{PC}+f^{PDC}\right)\end{pmatrix}c
\end{split}\end{equation}\end{linenomath}
These equations give the steady state \textsuperscript{13}C label distribution in citrate in the extracellular space or incubation medium after supplementation with {[2,5-\textsuperscript{13}C\textsubscript{2}]}glucose or {[2-\textsuperscript{13}C]}pyruvate. The distribution depends on the secretion fraction $c$ and on the fractions of \textsuperscript{13}C carbon entering the Krebs cycle via the pyruvate carboxylase route, $f^{PC}$, and via the pyruvate dehydrogenase complex route, $f^{PDC}$, (with $f^{PDC}=1-f^{PC}$). The distribution of \textsuperscript{13}C carbons for $f^{PC}=0.2)$ is shown in Figure \ref{fig:3} plotted against secretion fraction $c$.

%% file: p05_Model.tex
\subsection{CO\textsubscript{2} production}
Next to citrate production and secretion we also have to take into account how much carbon dioxide is produced every cycle and sum these contributions to see how much \textsuperscript{13}C labels are lost as carbon dioxide. During a Krebs cycle turn one six-carbon molecule of citrate is converted into a four-carbon metabolite oxaloacetate and two carbons are lost as carbon dioxide. The first carbon is lost during the conversion of isocitrate into $\alpha$-ketoglutarate and the second carbon is lost in the next step of the Krebs cycle during the conversion of $\alpha$-ketoglutarate into succinyl-CoA. The first originates from the C6 carbon in citrate and the latter originates from citrate C5. So, the carbon dioxide production at cycle turn $n$ is equal to the sum of citrate carbons C5 and C6 at this cycle turn $n$ times $(1-c)$ instead of times $c$ since the production of carbon dioxide during cycle $n$ takes place after the secretion of $(1-c)^n$ citrate. In other words, only the \textsuperscript{13}C carbons at C5 and C6, which are not secreted (fraction $(1-c)^n$) at cycle $n$, are lost as carbon dioxide. This means we could calculate the carbon dioxide production at cycle turn $n$ by summing the \textsuperscript{13}C distribution at C5 and C6 and multiply this by $\frac{1-c}{c}$. Because the \textsuperscript{13}C distribution at C5 and C6 is equal, this could be written as:
\begin{linenomath}\begin{equation}
    \overrightarrow{CO_2}(n)= 2\frac{1-c}{c}\overrightarrow{C5}(n)
\end{equation}\end{linenomath}
To calculate the resulting CO\textsubscript{2} produced over time the contributions of all cycle turns have to be summed.
\begin{linenomath}\begin{equation}
    \overrightarrow{CO_2}_{distr}=\sum\limits_{n=0}^{n=\infty}\overrightarrow{CO_2}(n)=2\frac{1-c}{c}\sum\limits_{n=0}^{n=\infty}\overrightarrow{C5}(n)
\end{equation}\end{linenomath}
Another way of calculating this is by evaluating the individual contributions for every cycle turn $\overrightarrow{CO_2}(n)$ to the total CO\textsubscript{2} for different values of $n$, like we did before for the carbons in citrate. For {[1,6-\textsuperscript{13}C\textsubscript{2}]}glucose or {[3-\textsuperscript{13}C]}pyruvate this looks like:
\begin{linenomath}\begin{align}
    \overrightarrow{CO_2}(n=0)&=0\\
    \overrightarrow{CO_2}(n=1)&=0\\
    \overrightarrow{CO_2}(n=2)&=\frac{1}{2}f^{PC}(1-c)^2\\
    \overrightarrow{CO_2}(n\geq3)&=cf^{PDC}\sum\limits^{i=n-2}_{i=1}\frac{(1-c)^n}{2^i}+\frac{(1-c)^n}{2^{n-1}}f^{PC}
\end{align}\end{linenomath}
This means we can calculate the total amount of CO\textsubscript{2} by summing these contributions:
\begin{linenomath}\begin{align}
    \overrightarrow{CO_2}_{distr}=&f^{PC}\sum\limits^{n=\infty}_{n=2}\frac{(1-c)^n}{2^{n-1}}+cf^{PDC}\sum\limits^{n=\infty}_{n=3}\sum\limits^{i=n-2}_{i=1}\frac{(1-c)^n}{2^i}\\
    =&(1-c)^3f^{PDC}-\frac{c(1-c)^3}{1+c}f^{PDC}+\frac{(1-c)^2}{1+c}f^{PC}\\
    =&\frac{2(1-c)}{c}\left[\frac{1-c}{2(1+c)}\left((1-c)f^{PDC}+f^{PC}\right)c\right]
\end{align}\end{linenomath}
In the last step the equation has been rewritten to show that the total CO\textsubscript{2} indeed is equal to the final \textsuperscript{13}C distribution at citrate C5 (or C6) times $2\frac{1-c}{c}$. Since the total \textsuperscript{13}C entering the Krebs cycle in our model is equal to $f^{PC}+f^{PDC} = 1$ and we only take citrate production/secretion and CO\textsubscript{2} into account in our model the total \textsuperscript{13}C found in citrate and CO\textsubscript{2} should also add up to 1. If we take the sum of the label distribution of the carbons of citrate and add the amount of CO\textsubscript{2} we should end up with $f^{PC}+f^{PDC}=1$. For {[1,6-\textsuperscript{13}C\textsubscript{2}]}glucose or {[3-\textsuperscript{13}C]}pyruvate this looks like:
\begin{linenomath}\begin{align}
    \overrightarrow{Citr}^{glc\mbox{-}1,6,sum}_{extr.cell.}&=cf^{PDC}\left(1+2\frac{1-c}{1+c}+\frac{(1-c)^2}{1+c}\right) + f^{PDC}\frac{(1-c)^3}{1+c}  \\ & +cf^{PC}\left(\frac{2}{1+c}+\frac{1-c}{1+c}\right) + f^{PC}\frac{(1-c)^2}{1+c}\\
    &=f^{PDC}+f^{PC}=1
\end{align}\end{linenomath}
We can repeat this exercise for the addition of {[2,5-\textsuperscript{13}C\textsubscript{2}]}glucose or {[2-\textsuperscript{13}C]}pyruvate. For this, we first have to take into account the values of $\overrightarrow{CO_2}(n)$ for different values of $n$, like we did before for the carbons in citrate:
\begin{linenomath}\begin{align}
    \overrightarrow{CO_2}(n=0)&=0\\
    \overrightarrow{CO_2}(n=1)&=0\\
    \overrightarrow{CO_2}(n\geq2)&=(cf^{PDC}+\frac{1}{2^{n-1}}f^{PC})(1-c)^n
\end{align}\end{linenomath}
Now we can again calculate the total amount of CO\textsubscript{2} produced during the experiment by summing over $n$:
\begin{linenomath}\begin{align}
    \overrightarrow{CO_2}_{distr}=&cf^{PDC}\sum\limits^{n=\infty}_{n=2}(1-c)^n+f^{PC}\sum\limits^{n=\infty}_{n=2}\frac{(1-c)^n}{2^{n-1}}\\
    =&(1-c)^2f^{PDC}+\frac{(1-c)^2}{1+c}f^{PC}\\
    =&2\frac{(1-c)}{c}\left(\frac{1-c}{2}\left(\frac{1}{1+c}f^{PC}+f^{PDC}\right)c\right)
\end{align}\end{linenomath}
So again the CO\textsubscript{2} production is equal to the sum of carbon labeling at citrate C5 and C6 times the factor $\frac{1-c}{c}$. The total amount of carbon entering the system has to be equal to the amount of carbon ending up in either citrate carbons or CO\textsubscript{2}. If we take the sum of the labeling in the carbons of citrate and add the amount of CO\textsubscript{2} we should again end up with $f^{PC}+f^{PDC}=1$. For {[2,5-\textsuperscript{13}C\textsubscript{2}]}glucose or {[2-\textsuperscript{13}C]}pyruvate this looks like:
\begin{linenomath}\begin{align}
    \overrightarrow{Citr}^{glc\mbox{-}2,5,sum}_{extr.cell.}&=\left(\frac{2}{1+c}+\frac{1-c}{1+c}\right)cf^{PC}+\frac{(1-c)^2}{1+c}f^{PC}\\
    &+(2-c)cf^{PDC}+(1-c)^2f^{PDC}\\
    &=f^{PC}+f^{PDC}=1
\end{align}\end{linenomath}
So the total of \textsuperscript{13}C label entering the Krebs cycle is equal to the total \textsuperscript{13}C label we find after the experiment in citrate (extracellular or in the incubation medium) and in CO\textsubscript{2}.

\subsection{PDC versus PC contribution to citrate \textsuperscript{13}C-labeling}
The equations we derived for the \textsuperscript{13}C distribution over citrate carbons after \textsuperscript{13}C substrate application can easily be split in separate contributions of the pyruvate carboxylase route and the pyruvate dehydrogenase complex route. For the application of {[1,6-\textsuperscript{13}C\textsubscript{2}]}glucose or {[3-\textsuperscript{13}C]}pyruvate we arrive at the following expressions: \begin{linenomath}\begin{equation}\overrightarrow{^{13}C}^{glc\mbox{-}1,6}_{distr}=\begin{pmatrix}0\\[5pt]f^{PDC}\\[5pt]
    \frac{1-c}{1+c}f^{PDC}+\frac{1}{1+c}f^{PC}\\[5pt]
    \frac{1-c}{1+c}f^{PDC}+\frac{1}{1+c}f^{PC}\\[5pt]
    \frac{1-c}{2(1+c)}\left((1-c)f^{PDC}+f^{PC}\right)\\[5pt]
    \frac{1-c}{2(1+c)}\left((1-c)f^{PDC}+f^{PC}\right)\\[5pt]
    \frac{(1-c)^3}{c(1+c)}f^{PDC}+\frac{(1-c)^2}{c(1+c)}f^{PC}\end{pmatrix}c
    =\begin{pmatrix}0\\[5pt]1\\[5pt]
    \frac{1-c}{1+c}\\[5pt]
    \frac{1-c}{1+c}\\[5pt]
    \frac{(1-c)^2}{2(1+c)}\\[5pt]
    \frac{(1-c)^2}{2(1+c)}\\[5pt]
    \frac{(1-c)^3}{c(1+c)}\\[5pt]\end{pmatrix}cf^{PDC}
    +\begin{pmatrix}0\\[5pt]0\\[5pt]
    \frac{1}{1+c}\\[5pt]
    \frac{1}{1+c}\\[5pt]
    \frac{1-c}{2(1+c)}\\[5pt]
    \frac{1-c}{2(1+c)}\\[5pt]
    \frac{(1-c)^2}{c(1+c)}\\[5pt]\end{pmatrix}cf^{PC}\label{eq:2}\end{equation}\end{linenomath}
Note here that carbon dioxide is now also included in the vector (last value). And for the application of {[2,5-\textsuperscript{13}C\textsubscript{2}]}glucose or {[2-\textsuperscript{13}C]}pyruvate we can do the same and end up with the following equations:
\begin{linenomath}\begin{equation}
    \overrightarrow{^{13}C}^{glc\mbox{-}2,5}_{distr}=\begin{pmatrix}f^{PDC}\\[5pt]0\\[5pt]
    \frac{1}{1+c}f^{PC}\\[5pt]
    \frac{1}{1+c}f^{PC}\\[5pt]
    \frac{1-c}{2}\left(\frac{1}{1+c}f^{PC}+f^{PDC}\right)\\[5pt]
    \frac{1-c}{2}\left(\frac{1}{1+c}f^{PC}+f^{PDC}\right)\\[5pt]
    \frac{(1-c)^2}{c}f^{PDC}+\frac{(1-c)^2}{c(1+c)}f^{PC}\end{pmatrix}c
    =\begin{pmatrix}1\\[5pt]0\\[5pt]
    0\\[5pt]
    0\\[5pt]
    \frac{1-c}{2}\\[5pt]
    \frac{1-c}{2}\\[5pt]
    \frac{(1-c)^2}{c}\end{pmatrix}cf^{PDC}
    +\begin{pmatrix}0\\[5pt]0\\[5pt]
    \frac{1}{1+c}\\[5pt]
    \frac{1}{1+c}\\[5pt]
    \frac{1-c}{2(1+c)}\\[5pt]
    \frac{1-c}{2(1+c)}\\[5pt]
    \frac{(1-c)^2}{c(1+c)}\end{pmatrix}cf^{PC}\label{eq:3}
\end{equation}\end{linenomath}

\subsection{Overestimation of secretion fraction $c$} \label{sec:8}
Next to citrate, other molecules are extracted from the Krebs cycle for other metabolic processes. This means that the fraction of citrate molecules remaining in the cycle will be smaller than $(1-c)$. Let's define an apparent citrate secretion fraction $d$ which includes both citrate and other molecules (citrate equivalents) leaving the Krebs via other metabolic processes, for example in reactions involving aspartate and glutamate. So $d\geq c$ or $1-d \leq 1-c$. Now we can recalculate the equations derived earlier, first for {[1,6-\textsuperscript{13}C\textsubscript{2}]}glucose:
\begin{linenomath}\begin{equation}\begin{split}
\overrightarrow{^{13}C}_{distr,app. secr.}^{glc\mbox{-}1,6}=\sum\limits_{n=0}^{n=\infty}\left(\frac{1-d}{1-c}\right)^n\overrightarrow{C}(n)=&\begin{pmatrix}0\\[5pt]
    df^{PDC}\sum\limits_{j=0}^{j=\infty}(1-d)^j\\[5pt]
    df^{PDC}\sum\limits_{j=1}^{j=\infty}\sum\limits_{i=1}^{i=j}\frac{(1-d)^j}{2^i}+f^{PC}\sum\limits_{j=0}^{j=\infty}\frac{(1-d)^j}{2^{j+1}}\\[5pt]
    df^{PDC}\sum\limits_{j=1}^{j=\infty}\sum\limits_{i=1}^{i=j}\frac{(1-d)^j}{2^i}+f^{PC}\sum\limits_{j=0}^{j=\infty}\frac{(1-d)^j}{2^{j+1}}\\[5pt]
    df^{PDC}\sum\limits_{j=2}^{j=\infty}\sum\limits_{i=2}^{i=j}\frac{(1-d)^j}{2^i}+f^{PC}\sum\limits_{j=0}^{j=\infty}\frac{(1-d)^j}{2^{j+1}}\\[5pt]
    df^{PDC}\sum\limits_{j=2}^{j=\infty}\sum\limits_{i=2}^{i=j}\frac{(1-d)^j}{2^i}+f^{PC}\sum\limits_{j=0}^{j=\infty}\frac{(1-d)^j}{2^{j+1}}\\[5pt]\end{pmatrix}c\\
    =&\begin{pmatrix}0\\[5pt]f^{PDC}\\[5pt]
    \frac{1-d}{1+d}f^{PDC}+\frac{1}{1+d}f^{PC}\\[5pt]
    \frac{1-d}{1+d}f^{PDC}+\frac{1}{1+d}f^{PC}\\[5pt]
    \frac{1-d}{2(1+d)}\left((1-d)f^{PDC}+f^{PC}\right)\\[5pt]
    \frac{1-d}{2(1+d)}\left((1-d)f^{PDC}+f^{PC}\right)\\[5pt]\end{pmatrix}c\label{eq:5}
\end{split}\end{equation}\end{linenomath}
For {[2,5-\textsuperscript{13}C\textsubscript{2}]}glucose or {[2-\textsuperscript{13}C]}pyruvate we can do the same:
\begin{linenomath}\begin{equation}\begin{split}\overrightarrow{^{13}C}_{distr,app. secr.}^{glc\mbox{-}2,5}=\sum\limits_{n=0}^{n=\infty}\left(\frac{1-d}{1-c}\right)^n\overrightarrow{C}(n)=&\begin{pmatrix}d\sum\limits_{n=0}^{n=\infty}(1-d)^nf^{PDC}\\[5pt]0\\
    \frac{1}{2}\sum\limits_{i=0}^{i=\infty}\left(\frac{1-d}{2}\right)^if^{PC}\\[5pt]
    \frac{1}{2}\sum\limits_{i=0}^{i=\infty}\left(\frac{1-d}{2}\right)^if^{PC}\\[5pt]
    \frac{d}{2}\sum\limits_{i=1}^{i=\infty}(1-d)^if^{PDC}+\frac{1}{2}\sum\limits_{i=1}^{i=\infty}\left(\frac{1-d}{2}\right)^if^{PC}\\[5pt]
    \frac{d}{2}\sum\limits_{i=1}^{i=\infty}(1-d)^if^{PDC}+\frac{1}{2}\sum\limits_{i=1}^{i=\infty}\left(\frac{1-d}{2}\right)^if^{PC}\end{pmatrix}c\\
=&\quad\begin{pmatrix}f^{PDC}\\[5pt]0\\[5pt]
    \frac{1}{1+d}f^{PC}\\[5pt]
    \frac{1}{1+d}f^{PC}\\[5pt]
    \frac{1-d}{2}\left(\frac{1}{1+d}f^{PC}+f^{PDC}\right)\\[5pt]
    \frac{1-d}{2}\left(\frac{1}{1+d}f^{PC}+f^{PDC}\right)\end{pmatrix}c\label{eq:6}
\end{split}\end{equation}\end{linenomath}
The resulting equations are similar to the equations obtained earlier, but $c$ is replaced by $d$ everywhere and then all equations are multiplied by $\frac{c}{d}$. Taking this into account in the curves in figures \ref{fig:2} and \ref{fig:3}, would mean that they have $d$ instead of $c$ on the x-axis, and are scaled with factor $c/d$.

\subsection{\textsuperscript{13}C-enrichment of pyruvate and acetyl-CoA pool} \label{sec:11}
Until now, we assumed the \textsuperscript{13}C-enrichment of the pyruvate and acetyl-CoA to be 100\%. This does not represent the real situation in cells or tissue, therefore we introduce enrichment fractions $\epsilon_p$ and $\epsilon_a$. All equations are multiplied with $\epsilon_p$ and the terms involving carbons flowing via the PDC route are multiplied additionally with $\epsilon_a$. Another potential factor to take into consideration is the fast exchange between citrate, isocitrate and $\alpha$-ketoglutarate. During this exchange the C6 of citrate is exchanged with free CO\textsubscript{2}, which potentially is unlabeled. If there's any \textsuperscript{13}C labeling at C6, this might be diluted by this exchange with unlabeled CO\textsubscript{2}. To take the effect of this exchange on the citrate C6 level into account, a dilution factor $\epsilon_c$ is introduced. For {[1,6-\textsuperscript{13}C\textsubscript{2}]}glucose or {[3-\textsuperscript{13}C]}pyruvate this comes down to:
\begin{linenomath}\begin{equation}\begin{split} \label{eq:7}
\overrightarrow{^{13}C}_{distr}^{glc\mbox{-}1,6}=&\begin{pmatrix}0\\[5pt]\epsilon_af^{PDC}\\[5pt]\frac{(1-d)\epsilon_a}{1+d}f^{PDC}+\frac{1}{1+d}f^{PC}\\[5pt]\frac{(1-d)\epsilon_a}{1+d}\epsilon_af^{PDC}+\frac{1}{1+d}f^{PC}\\[5pt]\frac{1-d}{2(1+d)}\left((1-d)\epsilon_af^{PDC}+f^{PC}\right)\\[5pt]\frac{(1-d)\epsilon_c}{2(1+d)}\left((1-d)\epsilon_af^{PDC}+f^{PC}\right)\\[5pt]\frac{(1-d)^3\epsilon_a}{d(1+d)}f^{PDC}+\frac{(1-d)^2}{d(1+d)}f^{PC}\end{pmatrix}\epsilon_pd\end{split}\end{equation}\end{linenomath}
And for {[2,5-\textsuperscript{13}C\textsubscript{2}]}glucose or {[2-\textsuperscript{13}C]}pyruvate this would be:
\begin{linenomath}\begin{equation}\begin{split} \label{eq:8}
\overrightarrow{^{13}C}_{distr}^{glc\mbox{-}2,5}=&\begin{pmatrix}\epsilon_af^{PDC}\\[5pt]0\\[5pt]\frac{1}{1+d}f^{PC}\\[5pt]\frac{1}{1+d}f^{PC}\\[5pt]\frac{1-d}{2}\left(\frac{1}{1+d}f^{PC}+\epsilon_af^{PDC}\right)\\[5pt]\frac{(1-d)\epsilon_c}{2}\left(\frac{1}{1+d}f^{PC}+\epsilon_af^{PDC}\right)\\[5pt]\frac{(1-d)^2\epsilon_a}{d}f^{PDC}+\frac{(1-d)^2}{d(1+d)}f^{PC}\end{pmatrix}\epsilon_pd\end{split}\end{equation}\end{linenomath}

\subsection{Mass balance} \label{sec:12}
The efflux of other metabolites confined in the apparent secretion fraction $d$ does not mean that there is actually a fraction $d$ of Krebs cycle metabolites lost every cycle, but rather shows how much \textsuperscript{13}C labeled citrate equivalents are either diverged from the cycle or exchanged for unlabeled metabolites. This means that even if there's low influx of carbons via pyruvate carboxylase ($f^{PC}$), there's still citrate secretion possible, as long as the Krebs cycle is supplied sufficiently with carbons from unlabeled molecules (like glutamine, glutamate and aspartate). The apparent secretion fraction $d$ therefore tells us something about both citrate diverging from the Krebs cycle, citrate equivalent diverging from the Krebs cycle and the exchange of labeled molecules with unlabeled molecules (e.g. in the $\alpha$-ketoglutarate and (rather large) glutamate pool).

%% file: p06_Calculations.tex
\section{Calculation of secretion fraction $c$ and PC fraction $f^{PC}$}

\subsection{Experimental ratios}
From the equations obtained in sections \textit{"\nameref{sec:9}"} and \textit{"\nameref{sec:5}"} (also summarized in Table \ref{tab:2}) we can define experimental ratios of \textsuperscript{13}C citrate signal integrals which can be easily obtained from \textsuperscript{13}C MR spectra. Since citrate C2 and C4 on one hand and C1 and C5 on the other are chemically equivalent, we only have to consider ratios composed of C1/5, C2/4, C3 and C6. We suggest the following experimental ratios of fitted \textsuperscript{13}C citrate integrals to be used because they consist of the carbons with the highest signal intensity (and consequently SNR), for most of the values of $c$: $R_1$ for {[1,6-\textsuperscript{13}C\textsubscript{2}]}glucose or {[3-\textsuperscript{13}C]}pyruvate experiments.
\begin{linenomath}\begin{align}
    R_1 = \frac{(C_2+C_4)-C_3}{(C_2+C_4)}
\end{align}\end{linenomath}
The reason for choosing this ratio is that, when looking at the equations derived for the \textsuperscript{13}C distribution over the citrate, C3 and C4 have equal signal intensities. By subtracting C3 from C2+C4 (one peak) we get C2, so $R_1$ reports on the fraction that C2 makes up of C2+C4.
For the {[2,5-\textsuperscript{13}C\textsubscript{2}]}glucose or {[2-\textsuperscript{13}C]}pyruvate experiments we suggest $R_2$.
\begin{linenomath}\begin{align}
        R_2 &= \frac{C_1+C_5}{C_3}
\end{align}\end{linenomath}
These ratios can then be used as experimental input values for the quantitative model we derived in this work describing citrate production and secretion in prostate epithelial tissue or cell lines. After providing prostate tissue or cells either with {[1,6-\textsuperscript{13}C\textsubscript{2}]}glucose or {[3-\textsuperscript{13}C]}pyruvate or with {[2,5-\textsuperscript{13}C\textsubscript{2}]}glucose or {[2-\textsuperscript{13}C]}pyruvate, the extracellular material, being extracellular fluid or incubation medium, can be analyzed using \textsuperscript{13}C NMR spectroscopy. After identification and quantification of the citrate resonance peaks the two ratios, defined above, can be calculated. These are then used to calculate secretion fraction $c$ and pyruvate carboxylase contribution $f^{PC}$ as described below.

\subsection{Real citrate secretion fraction $c$}
With ratio $R_1$ (from the {[1,6-\textsuperscript{13}C\textsubscript{2}]}glucose or {[3-\textsuperscript{13}C]}pyruvate experiments) and ratio $R_2$ (from the {[2,5-\textsuperscript{13}C\textsubscript{2}]}glucose or {[2-\textsuperscript{13}C]}pyruvate experiments) we can calculate the secretion fraction. First we rewrite the equation for $R_1$:
\begin{linenomath}\begin{equation}
    R_1 = \frac{C_{2/4}-C_3}{C_{2/4}}=\frac{C_2+C_4-C_3}{C_2+C_4}=\frac{f^{PDC}+\frac{1-c}{1+c}f^{PDC}+\frac{1}{1+c}f^{PC}-\frac{1-c}{1+c}f^{PDC}-\frac{1}{1+c}f^{PC}}{f^{PDC}+\frac{1-c}{1+c}f^{PDC}+\frac{1}{1+c}f^{PC}}
\end{equation}\end{linenomath}
From {[2,5-\textsuperscript{13}C\textsubscript{2}]}glucose or {[2-\textsuperscript{13}C]}pyruvate experiments we can calculate ratio $R_2$:
\begin{linenomath}\begin{align}
R_2 =& \frac{C_{1/5}}{C_{3}}=\frac{C_1+C_5}{C_3}=\frac{f^{PDC}+\frac{1-c}{2}\left(\frac{1}{1+c}f^{PC}+f^{PDC}\right)}{\frac{1}{1+c}f^{PC}}\end{align}\end{linenomath}
This can be simplified by extracting an equation for $f^{PDC}$ dependent on $c$ and $R_1$, 
\begin{linenomath}\begin{equation}f^{PDC}=\frac{R_1}{(1+c)-R_1}\end{equation}\end{linenomath}
and substituting this into the equation for $R_2$:
\begin{linenomath}\begin{align}
R_2 =\frac{\frac{R_1}{1+c-R_1}+\frac{1-c}{2}\left(\frac{1}{1+c}\left(1-\frac{R_1}{1+c-R_1}\right)+\frac{R_1}{1+c-R_1}\right)}{\frac{1}{1+c}\left(1-\frac{R_1}{1+c-R_1}\right)}\end{align}\end{linenomath}
This can now be solved for $c$ by rewriting this into a quadratic equation:
\begin{linenomath}\begin{equation}-(R_1+1)c^2 + (4R_1-2R_2)c + (R_1-2R_2+4R_1R_2+1)=0\end{equation}\end{linenomath}
After substituting values for $R_1$ and $R_2$ in this quadratic equation, we can solve for $c$. For example, for for $c=0.3$ and $f^{PC}=0.2$, $R_1$ would be $\approx0.58$ and $R_2$ would be $\approx7.37$, filling this in gives:
\begin{linenomath}\begin{equation}-1.58c^2 -12.43c + 3.87=0\end{equation}\end{linenomath}
Solving the quadratic equation of course gives us back $c=0.3$.

\subsection{Fraction $f^{PC}$ versus $f^{PDC}$} \label{sec:1}
\begin{figure}[!htb]
    \centering
    \includegraphics[width=1\textwidth]{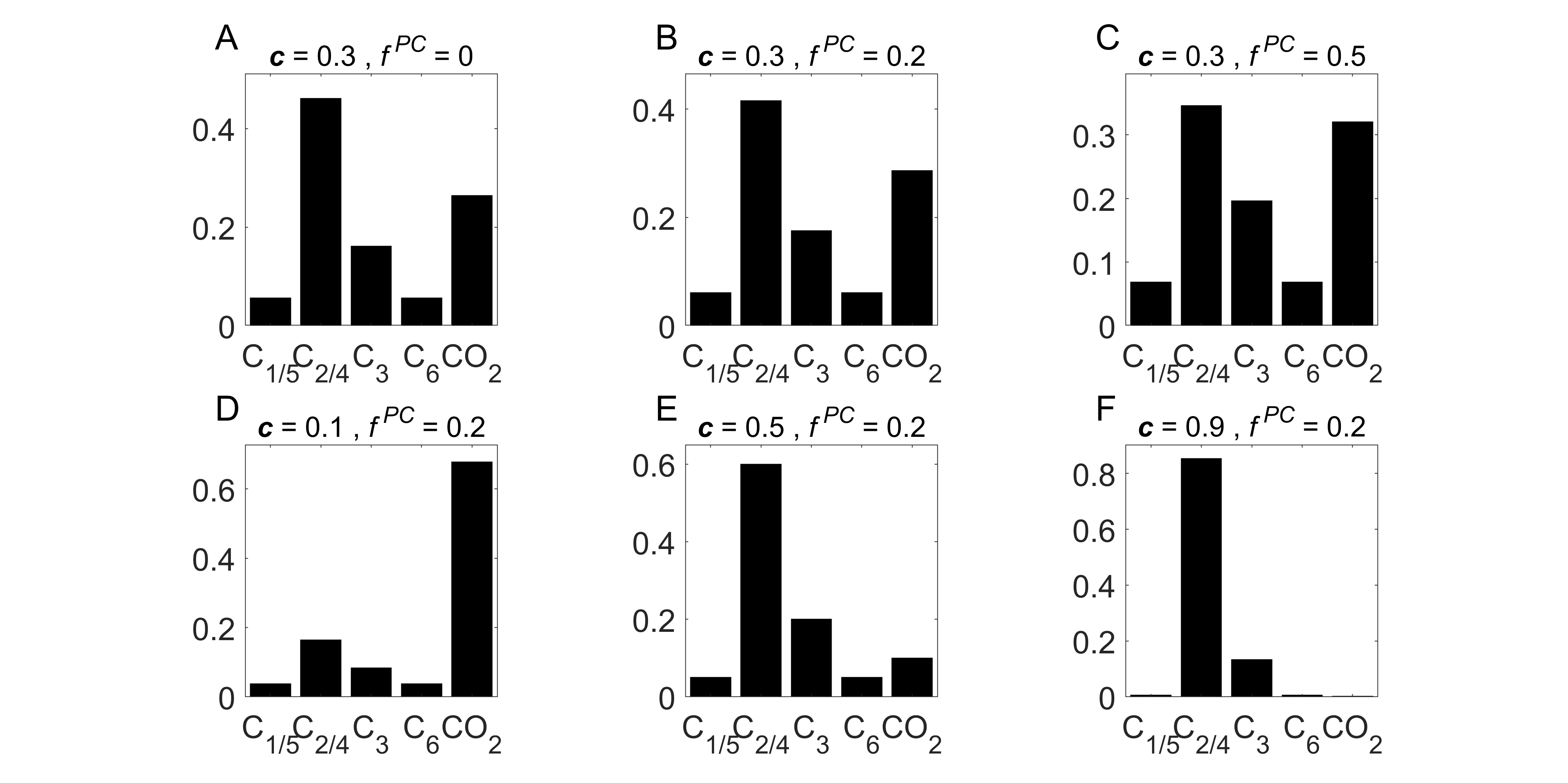}
    \caption{Relative \textsuperscript{13}C NMR signal integrals of the six citrate carbons and carbon dioxide for different values of $f^{PC}$ with $c=0.3$ (A-C), and different values of $c$ with $f^{PC}=0.2$ (D-F), after the application of {[1,6-\textsuperscript{13}C\textsubscript{2}]}glucose or {[3-\textsuperscript{13}C]}pyruvate to prostate epithelial tissue or cells. Carbons C1 and C5, and C2 and C4 have the same chemical shift and their signal integrals are therefore summed here. In these figures, the pyruvate and acetyl-CoA pool are assumed to be fully labeled and no other efflux of metabolites/citrate equivalents are taken into account.}
    \label{fig:4}
\end{figure}

\begin{figure}[!htb]
    \centering
    \includegraphics[width=1\textwidth]{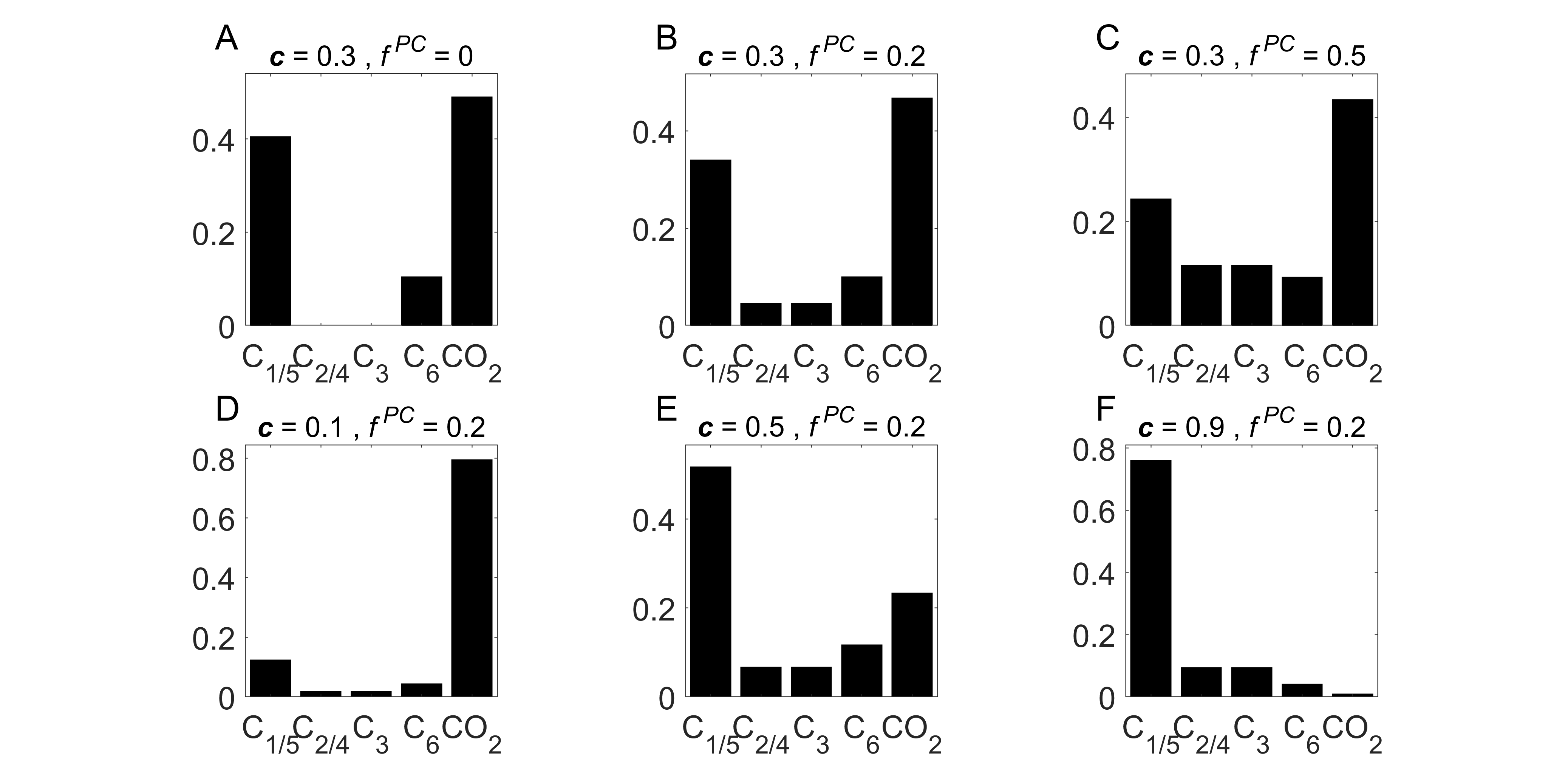}
    \caption{Relative \textsuperscript{13}C NMR signal integrals of the six citrate carbons and carbon dioxide for different values of $f^{PC}$ with $c=0.3$ (A-C), and different values of $c$ with $f^{PC}=0.2$ (D-F), after application of {[2,5-\textsuperscript{13}C\textsubscript{2}]}glucose or {[2-\textsuperscript{13}C]}pyruvate to prostate epithelial tissue or cells. Carbons C1 and C5, and C2 and C4 have the same chemical shift and their signal integrals are therefore summed here. In these figures, the pyruvate and acetyl-CoA pool are assumed to be fully labeled and no other efflux of metabolites/citrate equivalents are taken into account.}
    \label{fig:5}
\end{figure}

Now, using the expression we derived above we can calculate the fractions $f^{PDC}$ and $f^{PC}$:
\begin{linenomath}\begin{align}
    f^{PDC}&=\frac{R_1}{1+c-R_1}\\
    f^{PC}=&1-f^{PDC}\label{eq:4}
\end{align}\end{linenomath}
Using the values for $c$ and $f^{PC}$ we can calculate the total integrals of the citrate and carbon dioxide carbons according to equations \ref{eq:2} and \ref{eq:3} (see also Figures \ref{fig:4} and \ref{fig:5}). Looking at the predicted \textsuperscript{13}C integrals for the different citrate carbons assuming $c=0.3$ and $f^{PC}=0.2$, we can see that indeed C2/4 and C3 in the {[1,6-\textsuperscript{13}C\textsubscript{2}]}glucose or {[3-\textsuperscript{13}C]}pyruvate experiment is the best choice taking into account SNR, and for the {[2,5-\textsuperscript{13}C\textsubscript{2}]}glucose or {[2-\textsuperscript{13}C]}pyruvate experiment the C\textsubscript{1/5} and C\textsubscript{3} or C\textsubscript{6} signals would be best to evaluate. However, often the peak of citrate  C\textsubscript{6} overlaps with another peak in the \textsuperscript{13}C spectrum (e.g. pyroglutamate), making C\textsubscript{3} the best choice.

\subsection{Average number of Krebs cycle turns}
Using equation \ref{eq:1} derived in section \textit{"\nameref{sec:7}"} the average number of Krebs cycle turns that are completed before secretion of a citrate molecule can be calculated:
\begin{linenomath}\begin{equation}
    <n>=\sum\limits_{n=0}^{n=\infty}nc(1-c)^n=\frac{1-c}{c}
\end{equation}\end{linenomath}

\subsection{Pyruvate carboxylase fraction dependence on secretion fraction}
\begin{figure}[!htb]
    \centering
    \includegraphics[width=1\textwidth]{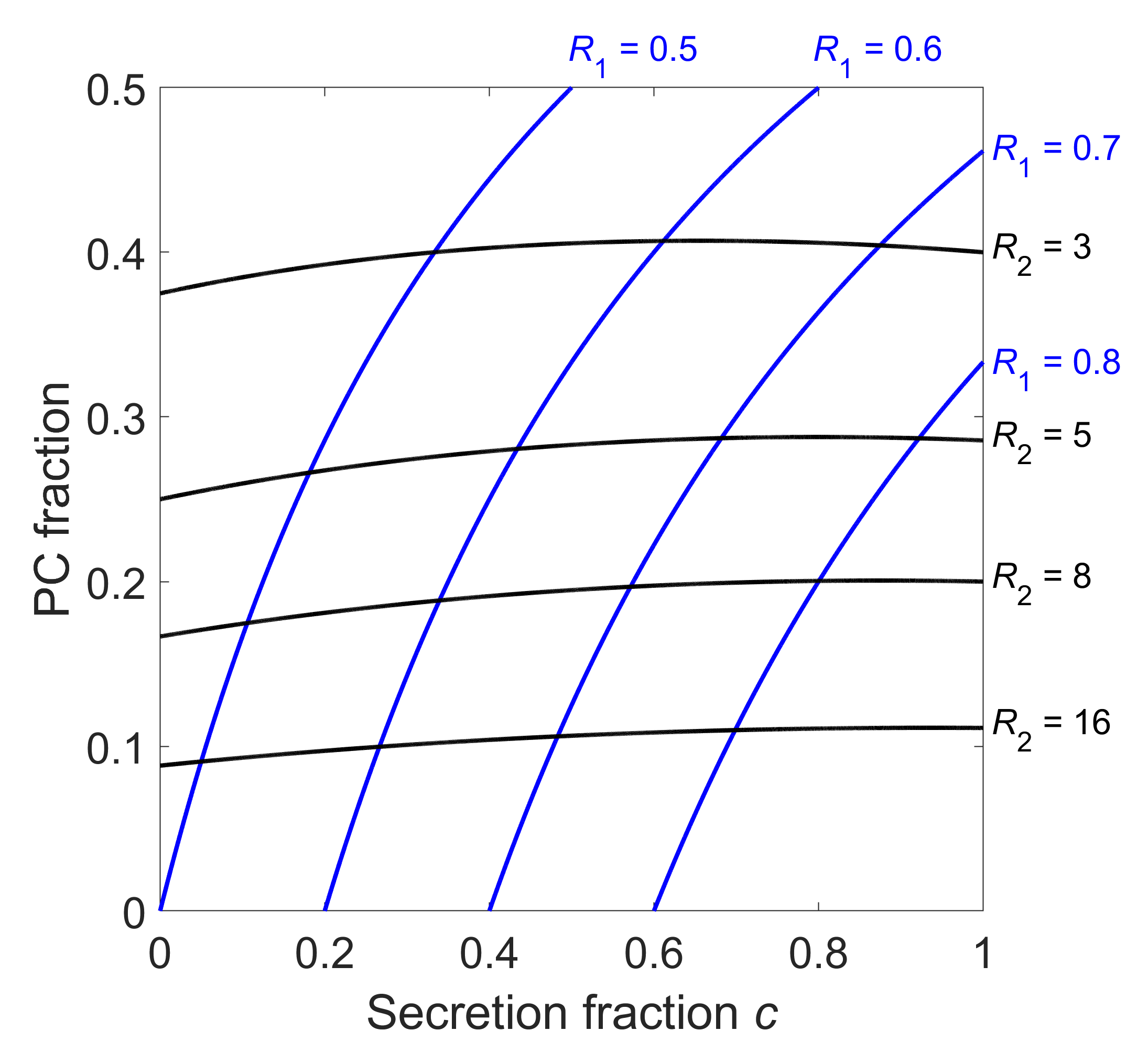}
    \caption{Fraction pyruvate carboxylase $f^{PC}$ versus secretion fraction $c$ with lines representing different values for $R_1$ and $R_2$.}
    \label{fig:7}
\end{figure}

In section \textit{"\nameref{sec:1}"} (vide supra) an expression was derived for fractions $f^{PDC}$ and $f^{PC}$ (using $f^{PC}=1-f^{PDC}$) depending on $c$ and $R_1$ (Equation \ref{eq:4}). We can derive a similar equation for $f^{PC}$ only depending on $c$ and $R_2$.
\begin{linenomath}\begin{align}
R_2 =& \frac{C_{1/5}}{C_{3}}=\frac{C_1+C_5}{C_3}=\frac{f^{PDC}+\frac{1-c}{2}\left(\frac{1}{1+c}f^{PC}+f^{PDC}\right)}{\frac{1}{1+c}f^{PC}}\\
f^{PC}=& \frac{(1+c)(3-c)}{2(R_2+1)+(3-c)c}
\end{align}\end{linenomath}
Using this and equation \ref{eq:4}, we can plot $c$ versus $f^{PC}$ for different values of $R_1$ and $R_2$ (Figure \ref{fig:7}). The intersection of the lines for $R_1$ and $R_2$ represent a unique combination of secretion fraction $c$ and pyruvate carboxylase fraction $f^{PC}$, derived from these ratios of \textsuperscript{13}C NMR signal intensities of citrate carbons.

\subsection{Error propagation}
\begin{figure}[!htb]
    \centering
    \includegraphics[width=1\textwidth]{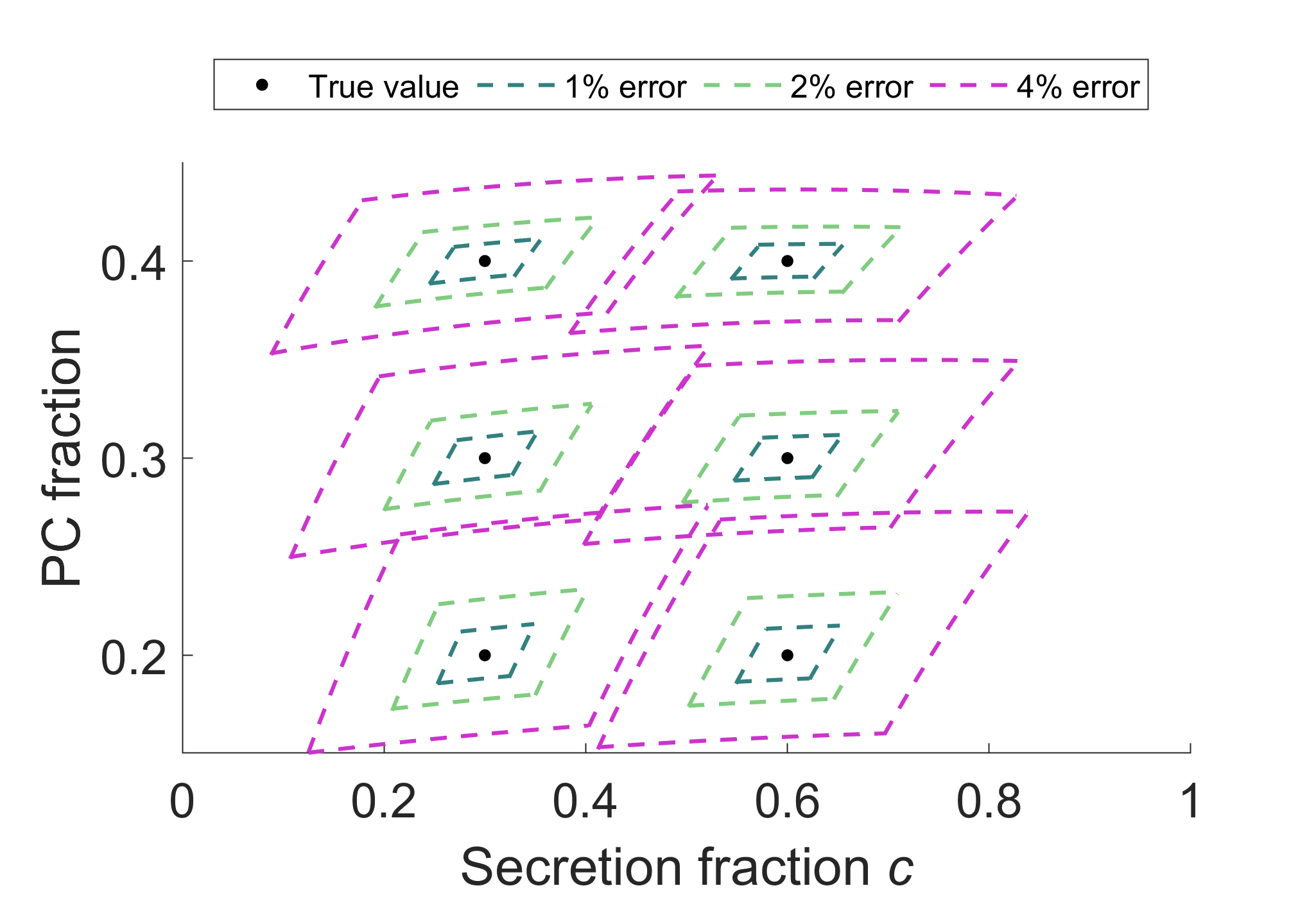}
    \caption{Confidence intervals for pyruvate carboxylase fraction $f^{PC}$ and secretion fraction $c$ at different error levels. These intervals are shown for $f^{PC}=0.2$, $f^{PC}=0.3$ and $f^{PC}=0.4$, and for the secretion fraction $c=0.3$ and $c=0.6$. Confidence intervals were calculated for errors (or noise levels) of 1\%, 2\% and 4\% of the biggest integral in the equations for $R_1$ and $R_2$ (From the equations derived in sections \textit{"\nameref{sec:5}"} and \textit{"\nameref{sec:9}"} follows that C\textsubscript{2/4} is the largest integral for $R_1$ and C\textsubscript{1/5} for $R_2$ for all physical values of $c$ and $f^{PC}$).}
    \label{fig:6}
\end{figure}

The experimental ratios obtained in \textsuperscript{13}C experiments are subject to errors due to SNR limitations of the spectra and this has its effect on variability of the values for $c$ and $f^{PC}$ that we calculated. Assuming a standard deviation equal to 1\%, 2\% or 4\% of the largest signal integral (e.g. C2) for all components in the ratio (in a way similar to relative noise levels of 1\%, 2\% and 4\%) we calculated the error propagation for $c$ and $f^{PC}$. For $c=0.3$ and $f^{PC}=0.2$ the error bounds are presented in figure \ref{fig:6}. A large spread in possible values for $c$ is seen at increasing error, while a more moderate spread is observed in values for $f^{PC}$.

\subsection{Efflux from the Krebs cycle of other metabolites in addition to citrate and different pool \textsuperscript{13}C-enrichments}
\subsubsection{Secretion fraction with efflux of other metabolites in addition to citrate}
In the previous sections we neglected the contributions of other metabolic pathways (both cataplerotic and anaplerotic), connected to the Krebs cycle carbon pool, in the efflux of carbons from this cycle. To take this into account we introduced an 'apparent citrate secretion fraction $d$' in the section \textit{"\nameref{sec:8}"} covering all contributions to the efflux of metabolites from the Krebs cycle. We also assumed equal (full) \textsuperscript{13}C-enrichment of the pyruvate and acetyl-CoA pools. We can use the results in Equations  \ref{eq:7} and \ref{eq:8} to recalculate the results and see what the effect of these parameters is. Using $R_1$ and $R_2$ and the equations in this section (Eq. \ref{eq:5} and \ref{eq:6}) we can only calculate the total fraction of molecules flowing out of the Krebs cycle, $d$. Within the current experimental scope we cannot separate secretion of citrate from other pathways transporting carbons out of the Krebs cycle. Still, we can assume $c\leq d$. An expression for $R_1$ including $d$ and the enrichment fractions is as follows:
\begin{linenomath}\begin{equation}
     R_1 = \frac{C_{2/4}-C_3}{C_{2/4}}=\frac{C_2+C_4-C_3}{C_2+C_4}=\frac{\epsilon_a f^{PDC}+\frac{1-d}{1+d}\epsilon_af^{PDC}+\frac{1}{1+d}f^{PC}-\frac{1-d}{1+d}\epsilon_af^{PDC}-\frac{1}{1+d}f^{PC}}{\epsilon_af^{PDC}+\frac{1-d}{1+d}\epsilon_af^{PDC}+\frac{1}{1+d}f^{PC}}\frac{\epsilon_p}{\epsilon_p}\\
\end{equation}\end{linenomath}
We can now again rewrite the expression to obtain $f^{PDC}$ as a function of $d$ and $R_1$:
\begin{linenomath}\begin{equation}f^{PDC}=\frac{R_1}{\epsilon_a(1+d)-R_1(2\epsilon_a-1)}\end{equation}\end{linenomath}
From the {[2,5-\textsuperscript{13}C\textsubscript{2}]}glucose or {[2-\textsuperscript{13}C]}pyruvate experiments we can calculate ratio $R_2$:
\begin{linenomath}\begin{align}
R_2 = \frac{C_{1/5}}{C_{3}}=\frac{C_1+C_5}{C_3}=\frac{\epsilon_af^{PDC}+\frac{1-d}{2}\left(\frac{1}{1+d}f^{PC}+\epsilon_af^{PDC}\right)}{\frac{1}{1+d}f^{PC}}\frac{\epsilon_p}{\epsilon_p}\end{align}\end{linenomath}
Substituting values for $R_1$ and $R_2$ we can now solve for $d$
\begin{linenomath}\begin{equation}-\epsilon_a(R_1+1)d^2 + \epsilon_a(4R_1-2R_2)d + \epsilon_a(R_1-2R_2+4R_1R_2+1)=0\end{equation}\end{linenomath}
Taking into account additional efflux of carbons from the Krebs cycle (Difference between apparent citrate secretion fraction $d$ and true citrate secretion fraction $c$), the real secretion fraction for citrate $c$ is evidently smaller or equal to the value found for $d$.

\subsubsection{Fraction $f^{PC}$ versus $f^{PDC}$ with efflux of other metabolites}
Above we presented an expression of fraction $f^{PDC}$), taking into account other efflux of citrate equivalents in an apparent secretion fraction $d$. We can derive an expression for $f^{PC}$:
\begin{linenomath}\begin{equation}
\begin{split}
    f^{PDC}=&\frac{R_1}{\epsilon_a(1+d)-R_1(2\epsilon_a-1)}\\
    f^{PC}=&1-f^{PDC}
\end{split}
\end{equation}\end{linenomath}
Taking into account other efflux besides citrate from the Krebs cycle, fortunately does not affect the values calculated for $f^{PC}$ or $f^{PDC}$.

\subsubsection{Average number of Krebs cycle turns including efflux of citrate equivalents/other molecules $d$}
An expression for the average number of completed cycle turns before secretion of a citrate molecule, taking an apparent secretion fraction $d$ into account, is straightforwardly obtained by replacing $c$ for $d$.
\begin{linenomath}\begin{equation}
    <n>=\sum\limits_{n=0}^{n=\infty}nd(1-d)^n=\frac{1-d}{d}
\end{equation}\end{linenomath}

\subsubsection{True citrate secretion fraction $c$ versus apparent citrate secretion fraction $d$}
It is not possible to distinguish $c$ and $d$ from each other within the experimental scope for which our model is designed, but we do know that the secretion fraction $c$ has to be smaller than $d$ if there is carbon, in addition to citrate carbons, flowing out of the Krebs cycle. So less than the calculated fraction $d$ of citrate molecules are secreted every cycle turn. The average number of completed cycle turns before secretion of citrate, in the situation that there are no other processes involved, would be even larger.

%% file: p10_Table_2.tex
\begin{landscape}
\thispagestyle{empty}
\begin{table}[ht]
\centering
\makebox[\linewidth]{
$\begin{array}{lcccccccccccc}
\overrightarrow{^{13}C}_{distr}
 &=&\overrightarrow{C}(0)
 &+\overrightarrow{C}(1)
 &+\overrightarrow{C}(2)
 &+\overrightarrow{C}(3)&+...
 &=\sum\limits_{n=0}^{n=\infty}\overrightarrow{C}(n)\\[8pt]\hline
\overrightarrow{^{13}C}_{distr}^{glc\mbox{-}1,6}
 &=&\begin{pmatrix}0\\[5pt]cf^{PDC}\\[5pt]\frac{1}{2}f^{PC}\\[5pt]\frac{1}{2}f^{PC}\\[5pt]0\\[5pt]0\\[5pt]0\\[5pt]\end{pmatrix}c
 &+\begin{pmatrix}0\\[5pt]cf^{PDC}\\[5pt]\frac{c}{2}f^{PDC}+\frac{1}{4}f^{PC}\\[5pt]\frac{c}{2}f^{PDC}+\frac{1}{4}f^{PC}\\[5pt]\frac{1}{4}f^{PC}\\[5pt]\frac{1}{4}f^{PC}\\[5pt]0\\[5pt]\end{pmatrix}c(1-c)
 &+\begin{pmatrix}0\\[5pt]cf^{PDC}\\[5pt]\frac{3c}{4}f^{PDC}+\frac{1}{8}f^{PC}\\[5pt]\frac{3c}{4}f^{PDC}+\frac{1}{8}f^{PC}\\[5pt]\frac{c}{4}f^{PDC}+\frac{1}{8}f^{PC}\\[5pt]\frac{c}{4}f^{PDC}+\frac{1}{8}f^{PC}\\[5pt]\frac{1}{2c}f^{PC}\\[5pt]\end{pmatrix}c(1-c)^2
 &+\begin{pmatrix}0\\[5pt]cf^{PDC}\\[5pt]\frac{7c}{8}f^{PDC}+\frac{1}{16}f^{PC}\\[5pt]\frac{7c}{8}f^{PDC}+\frac{1}{16}f^{PC}\\[5pt]\frac{3c}{8}f^{PDC}+\frac{1}{16}f^{PC}\\[5pt]\frac{3c}{8}f^{PDC}+\frac{1}{16}f^{PC}\\[5pt]\frac{1}{2}f^{PDC}+\frac{1}{4c}f^{PC}\\[5pt]\end{pmatrix}c(1-c)^3&+...
 &=\begin{pmatrix}0\\[5pt]f^{PDC}\\[5pt]
    \frac{1-c}{1+c}f^{PDC}+\frac{1}{1+c}f^{PC}\\[5pt]
    \frac{1-c}{1+c}f^{PDC}+\frac{1}{1+c}f^{PC}\\[5pt]
    \frac{1-c}{2(1+c)}\left((1-c)f^{PDC}+f^{PC}\right)\\[5pt]
    \frac{1-c}{2(1+c)}\left((1-c)f^{PDC}+f^{PC}\right)\\[5pt]
    \frac{(1-c)^3}{c(1+c)}f^{PDC}+\frac{(1-c)^2}{c(1+c)}f^{PC}\end{pmatrix}c\\
\overrightarrow{^{13}C}_{distr}^{glc\mbox{-}2,5}
 &=&\begin{pmatrix}cf^{PDC}\\[5pt]0\\[5pt]\frac{1}{2}f^{PC}\\[5pt]\frac{1}{2}f^{PC}\\[5pt]0\\[5pt]0\\[5pt]0\\[5pt]\end{pmatrix}c
 &+\begin{pmatrix}cf^{PDC}\\[5pt]0\\[5pt]\frac{1}{4}f^{PC}\\[5pt]\frac{1}{4}f^{PC}\\[5pt]\frac{c}{2}f^{PDC}+\frac{1}{4}f^{PC}\\[5pt]\frac{c}{2}f^{PDC}+\frac{1}{4}f^{PC}\\[5pt]0\\[5pt]\end{pmatrix}c(1-c)
 &+\begin{pmatrix}cf^{PDC}\\[5pt]0\\[5pt]\frac{1}{8}f^{PC}\\[5pt]\frac{1}{8}f^{PC}\\[5pt]\frac{c}{2}f^{PDC}+\frac{1}{8}f^{PC}\\[5pt]\frac{c}{2}f^{PDC}+\frac{1}{8}f^{PC}\\[5pt]f^{PDC}+\frac{1}{2c}f^{PC}\\[5pt]\end{pmatrix}c(1-c)^2
 &+\begin{pmatrix}cf^{PDC}\\[5pt]0\\[5pt]\frac{1}{16}f^{PC}\\[5pt]\frac{1}{16}f^{PC}\\[5pt]\frac{c}{2}f^{PDC}+\frac{1}{16}f^{PC}\\[5pt]\frac{c}{2}f^{PDC}+\frac{1}{16}f^{PC}\\[5pt]f^{PDC}+\frac{1}{4c}f^{PC}\\[5pt]\end{pmatrix}c(1-c)^3&+...
 &=\begin{pmatrix}f^{PDC}\\[5pt]0\\[5pt]
    \frac{1}{1+c}f^{PC}\\[5pt]
    \frac{1}{1+c}f^{PC}\\[5pt]
    \frac{1-c}{2}\left(\frac{1}{1+c}f^{PC}+f^{PDC}\right)\\[5pt]
    \frac{1-c}{2}\left(\frac{1}{1+c}f^{PC}+f^{PDC}\right)\\[5pt]
    \frac{(1-c)^2}{c}f^{PDC}+\frac{(1-c)^2}{c(1+c)}f^{PC}\end{pmatrix}c
\end{array}$}
\caption{The \textsuperscript{13}C distributions over six citrate carbons and carbon dioxide for the Krebs cycle turns $n=0-3$ and their summation. Top row is after providing prostate epithelial cells or tissue with {[1,6-\textsuperscript{13}C\textsubscript{2}]}glucose or {[3-\textsuperscript{13}C]}pyruvate
and bottom row shows the result after providing {[2,5-\textsuperscript{13}C\textsubscript{2}]}glucose or {[2-\textsuperscript{13}C]}pyruvate.}
\label{tab:2}
\end{table}
\end{landscape}
\clearpage

%% file: p07_DiscussionConclusion.tex
\section{Discussion \& Conclusions}
In this paper we present a quantitative model describing metabolic pathways in the production of labeled citrate in the mitochondria of prostatic epithelial cells and its secretion into the luminal space or incubation medium. Our model focuses on the application of \textsuperscript{13}C labeled substrates and the distribution of \textsuperscript{13}C labels over the 6 carbons in secreted citrate as a read-out for intracellular metabolism. In the design of the model we have taken common characteristics of the Krebs cycle into account such as the rapid carbon exchange between oxaloacetate, malate and fumarate.\supercite{Magnusson1991,Merritt2011} We show how to calculate the fraction of citrate leaving the Krebs cycle for secretion, i.e. $c$, and how to calculate the fractions of pyruvate dehydrogenase complex (PDC) versus the anaplerotic pyruvate carboxylase (PC) pathway in supplying the Krebs cycle with carbons from pyruvate, i.e. $f^{PC}$ and $f^{PDC}$. For this purpose we suggest two simple ratios of \textsuperscript{13}C NMR signal integrals of citrate that can be obtained after supplementation with \textsuperscript{13}C labeled glucose or pyruvate to prostate tissue \textit{in vivo} or cell lines \textit{in vitro}. These measures are shown to be independent of \textsuperscript{13}C-enrichment of the administered supplements. Since the total NMR integrals of the citrate carbons are measured, the model is independent of \textsuperscript{13}C J-coupling patterns, making this method more robust if SNR is low. It can be adjusted for use with other \textsuperscript{13}C labeled supplements, especially if those are feeding the acetyl-CoA pool.
\bigskip\\
For this model we assumed that the \textsuperscript{13}C pool of glucose and pyruvate are large compared to the Krebs cycle \textsuperscript{13}C pool, which results in an average constant flow of \textsuperscript{13}C labels entering the Krebs cycle over time. Another assumption made in deriving this model is the constant secretion fraction of citrate over time. If the tissue or cells are provided with a constant flow of \textsuperscript{13}C label for a long period of time, we can assume that the epithelial cells on average are in a steady state with respect to citrate production and general metabolism during this period.
\bigskip\\
After citrate diverges from the Krebs cycle and is transported to the cytosol it can be used in \textit{de novo} lipid sythesis and in further metabolic conversions. Assuming lipid synthesis and secretion of citrate by the transporters out of the cells is constant over time this doesn't influence the \textsuperscript{13}C labeling pattern found for citrate in the extracellular fluid or incubation medium since an equal fraction of all citrate present in the cytosol will be lost to lipid synthesis or other metabolism independent of the \textsuperscript{13}C labeling of citrate. Only if the \textsuperscript{13}C labels subsequently end up in the Krebs cycle again this could influence the final outcome of the experiments, but we assume this to be negligible.
\bigskip\\
Initially we assumed that, apart from secretion of citrate and carbon dioxide, no other efflux of carbons occurs during each cycle turn. In this way we derived an expression for the average true citrate secretion fraction $c$. However, carbons may leave the Krebs cycle elsewhere. Thus, the secretion fraction calculated from experimental values using this model concerns the total fraction of molecules diverging from the Krebs cycle during one cycle turn, not only citrate. We show that the efflux of other metabolites does not have an effect on the value of $f^{PC}$ or $f^{PDC}$ but it does result in an overestimation of secretion fraction $c$.To take this into account we introduce an apparent secretion fraction $d$, which represents Krebs cycle efflux of carbons from citrate and other molecules and thus gives an upper value for the secretion of citrate. Efflux from the Krebs cycle pool can be due to for example conversion of $\alpha$-ketoglutarate into glutamate by glutamate dehydrogenase, pyruvate production from oxaloacetate via PEPCK and aspartate transaminase activity. The latter is part of the malate-aspartate shuttle and catalyzes the conversion of $\alpha$-ketoglutarate into glutamate and of oxaloacetate into aspartate, and contributes to the efflux of metabolites from the Krebs cycle for production of amino acids and proteins. Multiple metabolic flux modeling studies have calculated total Krebs cycle fluxes and net influx and efflux from the Krebs cycle. In some cases exchange fluxes were calculated in addition to net fluxes. Especially interesting is the high exchange flux associated with the exchange between $\alpha$-ketoglutarate and glutamate, which can be comparable to total Krebs cycle fluxes \textit{in vivo} in tissues like heart\supercite{Burgess2001,Carvalho2004} and brain\supercite{Chen2001,Mason1995,Xin2015,Yang2006}, but also \textit{in vitro} in metastatic cell lines like melanoma\supercite{Shestov2016,Shestov2016a}, hepatocyte and hepatoma\supercite{Egnatchik2018} and glioma\supercite{PORTAIS1993} cells. Depending on the size of the unlabeled glutamate pool this can result in a fast isotopical dilution of the $\alpha$-ketoglutarate pool, resulting in loss of \textsuperscript{13}C labeled carbon skeletons. These processes and possibly further loss of \textsuperscript{13}C-labeled Krebs cycle metabolites accounts for the total efflux of \textsuperscript{13}C carbons from the Krebs cycle. Until now no literature values for similar exchange fluxes are available for prostate tissues or cell systems. Future efforts can be made to extend the model with an estimation of loss of \textsuperscript{13}C labeled carbon skeletons to glutamate using the intracellular glutamate pool size and (relative) exchange fluxes. The \textsuperscript{13}C-enrichment of the pyruvate and acetyl-CoA pool are taken into consideration as well, assuming these to be equal leads to an overestimation of the fraction $f^{PC}$. Therefore it is needed to include analysis of enrichments of these pools into experimental setups.
\bigskip\\
The high labeling efficiency of citrate using the \textsuperscript{13}C substrates described here make these substrates ideal candidates for studying metabolism underlying the unique secretion of citrate by prostate tissue. Still, extending this model to other \textsuperscript{13}C labeled substrates or other isotopes like \textsuperscript{14}C or \textsuperscript{2}H would open windows on more details of this metabolism.
\bigskip\\
In the context of our model we propose ratios of \textsuperscript{13}C NMR signals of citrate, $R_1$ and $R_2$, as read-out indices of metabolism involved in citrate production and secretion in prostatic tissue. From these indices we can calculate Krebs cycle efflux or true citrate secretion fraction $c$ (or apparent citrate secretion fraction $d$) and the fraction of pyruvate carbons entering the Krebs cycle via the PC or PDC routes. A change in these indices could indicate a shift in metabolism associated with the development of prostate diseases, such as prostate cancer, and could be used to probe therapy effectiveness. For this it would be necessary to first measure baseline indices in healthy prostate tissue by measuring citrate secreted by healthy epithelial tissue \textit{in vivo} and secondly determine how much these indices changes upon disease, in particular malignancy, or treatment.

%% file: p08_Appendices.tex
\section{Appendix}
\subsection{Propagation of error}
The error in estimating the true integrals of citrate \textsuperscript{13}C carbons found in \textsuperscript{13}C MR spectra results in an error in the estimation of proposed ratios $R_1$ and $R_2$.
\begin{linenomath}\begin{align}
    R_1 &= \frac{C2/4-C3}{C2/4}\\
    R_1 \pm \sigma_{R_1}&= \frac{(C2/4\pm\sigma_{C2/4})-(C3\pm\sigma_{C3})}{(C2/4\pm\sigma_{C2/4})}\\
\end{align}\end{linenomath}
We assume a similar error on all \textsuperscript{13}C MR signals, so $\sigma_{C3}=\sigma_{C2/4}=\sigma$. This means the standard error can be estimated using the following equations:
\begin{linenomath}\begin{align}
    \frac{\sigma_{R_1}}{R_1} &\approx \sqrt{\frac{2\sigma^2}{(C2/4-C3)^2}+\frac{\sigma^2}{C2/4^2}}\\
    &= \sqrt{\frac{2\sigma^2}{R_1^2C2/4^2}+\frac{\sigma^2}{C2/4^2}}\\
    &= \frac{\sigma}{C2/4}\sqrt{\frac{2}{R_1^2}+1}
\end{align}\end{linenomath}
This means that if we assume the standard error to be a fraction $\alpha$ of the largest integral, there are two possible solutions
\begin{linenomath}\begin{align}
    \frac{\sigma_{R_1}}{R_1} &\approx \alpha\sqrt{\frac{2}{R_1^2}+1}\qquad \text{for $C2/4 \geq C3$, $\sigma=\alpha C2/4$}\\
    \frac{\sigma_{R_1}}{R_1} &\approx \alpha(1-R_1)\sqrt{\frac{2}{R_1^2}+1}\qquad \text{for $C2/4 \leq C3$, $\sigma=\alpha C3$}
\end{align}\end{linenomath}
\\
For ratio $R_2$ a similar derivation can be used
\begin{linenomath}\begin{align}
    R_2 &= \frac{C1/5}{C3}\\
    R_2 \pm \sigma_{R_2}&= \frac{(C1/5\pm\sigma_{C1/5})}{(C3\pm\sigma_{C3})}
 \end{align}\end{linenomath}
We assume a similar error on all \textsuperscript{13}C MR signals, so $\sigma_{C3}=\sigma_{C1/5}=\sigma$. This means the standard error can be estimated using the following equations:
    \begin{linenomath}\begin{align}
    \frac{\sigma_{R_2}}{R_2} &\approx \sqrt{\frac{\sigma^2}{C1/5^2}+\frac{\sigma^2}{C3}}\\
    &= \sqrt{\frac{\sigma^2}{R_2^2C3^2}+\frac{\sigma^2}{C3^2}}\\
    &= \frac{\sigma}{C3}\sqrt{\frac{1}{R_2^2}+1}
    \end{align}\end{linenomath}
This means that if we assume again that the standard error is a fraction $\alpha$ of the largest integral, there are two possible solutions:
    \begin{linenomath}\begin{align}
    \frac{\sigma_{R_2}}{R_2} &\approx \alpha\sqrt{\frac{1}{R_2^2}+1}\qquad \text{for $C3 \geq C1/5$, $\sigma=\alpha C3$}\\
    \frac{\sigma_{R_2}}{R_2} &\approx \alpha\sqrt{1+R_2^2}\qquad \text{for $C3 \leq C1/5$, $\sigma=\alpha C1/5$}
\end{align}\end{linenomath}
These equations are used in plotting the error bounds in Figure \ref{fig:6}.

\subsection{Average number of Krebs cycle turns for a \textsuperscript{13}C label}
\begin{figure}[!htb]
    \centering
    \includegraphics[width=1\textwidth]{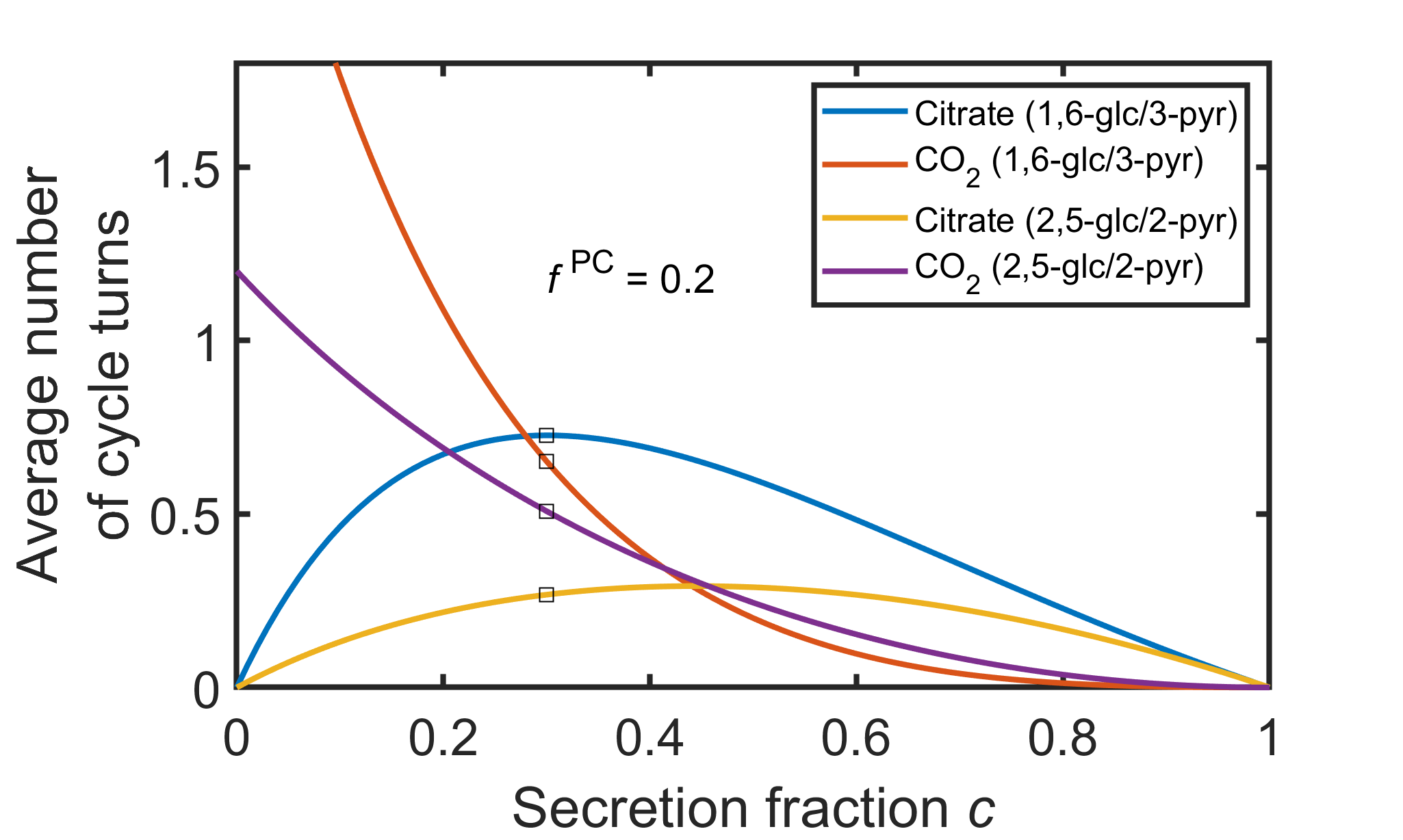}
    \caption{Average number of Krebs cycle turns a \textsuperscript{13}C carbon completes before being secreted in citrate or being lost into carbon dioxide versus secretion fraction $c$. The fraction pyruvate carboxylase is set to $f^{PC}=0.2$ and the values at $c=0.3$ are indicated with squares.}
    \label{fig:12}
\end{figure}

The average number of Krebs cycle turns that a \textsuperscript{13}C carbon completes before ending up in either citrate or carbon dioxide depends on the carbon position it starts out from at $n=0$. For a \textsuperscript{13}C carbon starting at citrate C2 or C3 the fraction of \textsuperscript{13}C carbon ending up in citrate during Krebs cycle turn $n$ is:
\begin{linenomath}\begin{align}
    C_{2/3}^{Citr}(n=0)&= c\\
    C_{2/3}^{Citr}(n=1)&= c(1-c)\\
    C_{2/3}^{Citr}(n\geq2)&= \frac{c(1-c)^n}{2^{n-2}}
\end{align}\end{linenomath}
And the fractions of \textsuperscript{13}C carbon ending up in carbon dioxide at cycle turn $n$ is:
\begin{linenomath}\begin{align}
    C_{2/3}^{CO_2}(n=0)&= 0\\
    C_{2/3}^{CO_2}(n=1)&= 0\\
    C_{2/3}^{CO_2}(n\geq2)&=\frac{(1-c)^{n+1}}{2^{n-1}}
\end{align}\end{linenomath}
Multiplying this by the corresponding Krebs cycle turn index $n$ and summing this over all possible values of $n$ gives us the average number of Krebs cycle turns a \textsuperscript{13}C carbon starting at C2 or C3 completes before ending up in citrate or carbon dioxide.
\begin{linenomath}\begin{align}
    <n_{C2/3}^{Citr}>&=\sum\limits_{n=0}^{n=\infty}nC_{2/3}^{Citr}(n)=c(1-c)\left(1+2\frac{(1-c)(3+c)}{(1+c)^2}\right)\\
    <n_{C2/3}^{CO_2}>&=\sum\limits_{n=0}^{n=\infty}nC_{2/3}^{CO_2}(n)=\frac{(3+c)(1-c)^3}{(1+c)^2}
\end{align}\end{linenomath}
We can do the same for citrate C1 and C4, this results in:
\begin{linenomath}\begin{align}
    C_{1/4}^{Citr}(n=0)&= c\\
    C_{1/4}^{Citr}(n=1)&= c(1-c)\\
    C_{1/4}^{Citr}(n\geq2)&= 0
    \end{align}\end{linenomath}
And the following contributions for cycle turn $n$ for carbon dioxide:
\begin{linenomath}\begin{align}
    C_{1/4}^{CO_2}(n=0)&=0\\
    C_{1/4}^{CO_2}(n=1)&=(1-c)^2\\
    C_{1/4}^{CO_2}(n\geq2)&= 0
    \end{align}\end{linenomath}
Again we can calculate the average number of cycle turns a \textsuperscript{13}C carbon starting at citrate C1 or C4 completes before secretion in citrate or ending up in carbon dioxide.
\begin{linenomath}\begin{align}
    <n_{C1/4}^{Citr}>&=\sum\limits_{n=0}^{n=\infty}nC_{1/4}^{Citr}(n)=c(1-c)\\
    <n_{C1/4}^{CO_2}>&=\sum\limits_{n=0}^{n=\infty}nC_{2/3}^{CO_2}(n)=(1-c)^2
\end{align}\end{linenomath}
Depending on the \textsuperscript{13}C substrate that is used the \textsuperscript{13}C carbons in citrate start at different positions at $n=0$. For {[1,6-\textsuperscript{13}C\textsubscript{2}]}glucose or {[3-\textsuperscript{13}C]}pyruvate \textsuperscript{13}C carbon ends up at C2 via the PDC route and at C3 or C4 via the PC route. For {[2,5-\textsuperscript{13}C\textsubscript{2}]}glucose or {[2-\textsuperscript{13}C]}pyruvate the \textsuperscript{13}C carbons start at citrate C1 via PDC and at citrate C3 or C4 via PC. The result is that the true average number of Krebs cycle turns a \textsuperscript{13}C carbon completes before ending up in citrate or carbon dioxide also depends on the value of $f^{PC}$. Calculating this for the {[1,6-\textsuperscript{13}C\textsubscript{2}]}glucose or {[3-\textsuperscript{13}C]}pyruvate substrates:
\begin{linenomath}\begin{align}
    <n_{glc\mbox{-}1,6}^{Citr}>&=\left(f^{PDC}+\frac{1}{2}f^{PC}\right)<n_{C2/3}^{Citr}>+\frac{1}{2}f^{PC}<n_{C1/4}^{Citr}>\\
    &=\left(f^{PDC}+\frac{1}{2}f^{PC}\right)c(1-c)\left(1+2\frac{(1-c)(3+c)}{(1+c)^2}\right)+f^{PC}\frac{c(1-c)}{2}\\
    <n_{glc\mbox{-}1,6}^{CO_2}>&=\left(f^{PDC}+\frac{1}{2}f^{PC}\right)<n_{C2/3}^{CO_2}>+\frac{1}{2}f^{PC}<n_{C1/4}^{CO_2}>\\
    &=\left(f^{PDC}+\frac{1}{2}f^{PC}\right)\frac{(3+c)(1-c)^3}{(1+c)^2}+f^{PC}\frac{(1-c)^2}{2}
\end{align}\end{linenomath}
And for the {[2,5-\textsuperscript{13}C\textsubscript{2}]}glucose or {[2-\textsuperscript{13}C]}pyruvate substrate
\begin{linenomath}\begin{align}
    <n_{glc\mbox{-}1,6}^{Citr}>&=\frac{1}{2}f^{PC}<n_{C2/3}^{Citr}>+\left(f^{PDC}+\frac{1}{2}f^{PC}\right)<n_{C1/4}^{Citr}>\\
    &=\frac{1}{2}f^{PC}c(1-c)\left(1+2\frac{(1-c)(3+c)}{(1+c)^2}\right)+\left(f^{PDC}+\frac{1}{2}f^{PC}\right)c(1-c)\\
    <n_{glc\mbox{-}1,6}^{CO_2}>&=\frac{1}{2}f^{PC}<n_{C2/3}^{CO_2}>+\left(f^{PDC}+\frac{1}{2}f^{PC}\right)<n_{C1/4}^{CO_2}>\\
    &=\frac{1}{2}f^{PC}\frac{(3+c)(1-c)^3}{(1+c)^2}+\left(f^{PDC}+\frac{1}{2}f^{PC}\right)(1-c)^2
\end{align}\end{linenomath}
The average number of Krebs cycle turns a \textsuperscript{13}C carbon completes before being secreted in citrate or carbon dioxide is plotted for $f^{PC}=0.2$ in Figure \ref{fig:12}. The longer a \textsuperscript{13}C stays in the Krebs cycle, the bigger the fraction that will be secreted as carbon dioxide, but also as citrate, the figure shows this balance and the optima. Depending on which \textsuperscript{13}C substrates are used and the pyruvate carboxylase fraction $f^{PC}$, the maximum of the citrate curves shifts. At this maximum a \textsuperscript{13}C carbon stays in citrate the longest.

